# Quasi-two-layer finite-volume scheme for modeling shallow water flows over an arbitrary bed in the presence of external force

K. V. Karelsky[1], A. S. Petrosyan[1,2], A. G. Slavin[3]

[1] Space Research Institute of the Russian Academy of Sciences, 117997, Moscow, Profsoyuznaya st. 84/32, 117997, Moscow

[2] Moscow Institute of Physics and Technology (State University), 9, Institutskii per., Dolgoprudny 41700, Moscow Region, Russia

[3] Memorial University of Newfoundland, Newfoundland and Labrador, Canada

Email addresses: kkarelsk@iki.rssi.ru (K. V. Karelsky), apetrosy@iki.rssi.ru (A. S. Petrosyan), aslavin@mun.ca (A. G. Slavin).

Finite-volume numerical method for study shallow water flows over an arbitrary bed profile in the presence of external force is proposed. This method uses the quasi-two-layer model of hydrodynamic flows over a stepwise boundary with advanced consideration of the flow features near the step. A distinctive feature of the suggested model is a separation of a studied flow into two layers in calculating flow quantities near each step, and improving by this means approximation of depth-averaged solutions of the initial three-dimensional Euler equations. We are solving the shallow-water equations for one layer, introducing the fictitious lower layer only as an auxiliary structure in setting up the appropriate Riemann problems for the upper layer. Besides quasi-two-layer approach leads to appearance of additional terms in one-layer finite-difference representation of balance equations. These terms provide the mechanical work made by nonhomogeneous bed interacting with flow. A notable advantage of the proposed method is the consideration of the properties of the process of the waterfall, namely the fluid flow on the step in which the fluid does not wet part of the vertical wall of the step. The presence of dry zones in the vertical part of the step indicates violation of the conditions of hydrostatic flow. The quasi-two-layer approach determines the size of the dry zone of the vertical component of the step. Consequently it gives an opportunity to figure out the amount of flow kinetic energy dissipation on complex boundary. Numerical simulations are performed based on the proposed algorithm of various physical phenomena, such as a breakdown of the rectangular fluid column over an inclined plane, large-scale motion of fluid in the gravity field in the presence of Coriolis force over an mounted obstacle on underlying surface. Computations are made for two dimensional dam-break problem on slope precisely conform to laboratory experiments. Interaction of the Tsunami wave with the shore line including an obstacle has been simulated to demonstrate the effectiveness of the developed algorithm in domains including partly flooded and dry regions.

----------------------------------------

Corresponding Author:

Professor Arakel Petrosyan
Space Research Institute

Russian Academy Sciences
117997, Moscow, Profsoyuznaya st. 84/32
Tel: +7-495-3335478
Fax: +7-495-3333011
E.mail: apetrosy@iki.rssi.ru

## 1. Introduction

The shallow water equations are derived from the non-stationary three-dimensional Euler equations by averaging over a vertical coordinate and taking into consideration hydrostatic pressure distribution [49]. It is expected that solutions of depth-averaged equations have similar properties as depth-averaged solutions of initial fluid equations. The obtained equations are also rather complicated for constructing general analytical solutions because of their nonlinearity and bed complexity [33, 34, 35]. However, they can be successfully integrated numerically [10, 50, 53].

The main difficulty in numerical simulation of nonhomogenous shallow water equations consists in their nonlinearity and their non-divergence property determined by nonhomogeneity of the right-hand side of the momentum conservation equations due to bed complexity. The presence of a non-divergent term induces highly nonlinear effects caused by stepwise change of hydrodynamic quantities in the areas of its sharp change in addition to nonlinear phenomena due to hyperbolic structure of shallow watyer equations. The other problem consists in compatibility of solutions of traditional depth-averaged equations with depth-averaged solutions of initial Euler equations due to significant role of shallow flows dependences on vertical coordinate [28]. Numerical methods have been developed and effectively used in studies of shallow water flows on complex boundary when an external force effects are insignificant.

The main problem in numerical integration of hyperbolic balance equations is in developing finite-difference schemes satisfying the conditions of conservation of stable states. In the context of Saint-Venant's equations this means the equilibrium of resting water. The schemes which satisfy such properties are called well-balanced. Roe modified scheme was suggested to use for satisfying the well-balanced conditions in [7]. This idea has been further developed in papers [43, 51]. The turbulence phenomena and mass sources in problems comprising nonhomogeneous bed are considered in papers [14, 15], and the improved reconstruction in [12, 13]. It was suggested in [40, 41] to use either ordinary or modified Godunov-type schemes (Riemann-solvers). This approach is well-proved and this idea has been widely applied to solve the problems with various source terms. In [5, 37] the method in which bed effects are compensated by changing of the values of fluid depths was suggested. In [3, 9, 10, 44] was proposed the method that uses the scheme of hydrostatic reconstruction of the flows on cells interfaces. The phenomena appearing because of presence of the Coriolis force were simulated using this scheme, particularly the geostrophic adaptation phenomenon. The same phenomena were considered in paper [38] in which LeVeque method is used [40, 41]. The propagation of tsunami waves and flooding processes are simulated in [19, 20, 21]. In [54] the surface gradient method of interpreting source terms in shallow water equations is proposed. It is based on the accurate reconstruction of conservative variables on cell interfaces. The method of separating flow quantities which keeps exact balance between the flow gradients and source terms is proposed in paper [25].

The method of high-order accuracy and its improving reconstruction is proposed in [52]. The exact partial solutions of the problem including a step-wise boundary on a bed are presented in [1]. The exact partial solutions for the problem including nonhomogeneous bed are presented in [8]. Also we have to mention the works [2, 4, 6, 11, 18, 21, 22, 23, 24, 50, 55] which made a significant contribution to solution of problems of simulating hydrodynamic flows over a nonhomogeneous bed with source terms of various origin. The main difficulty arising in applying such methods is the lack of a unique analytical solution of the one-dimensional nonlinear Riemann problem for step bed [27]. However, in all aforementioned works this problem was successfully solved by introducing some additional assumptions.

The presence of external forces, bed nonhomogeneities often causes vertical nonhomogeneity of horizontal shallow flows [28] that changes the values of hydrodynamic quantities averaged over the depth. It is necessary to take into consideration the vertical structure of depth-averaged fluid flows near the peculiarities of the bed to describe the indicated effects more adequately in numerical solutions. Existence of well-developed and validated numerical methods makes especially attractable the reduction of a problem in the presence an external force to solution of the problem of shallow water flows over a composed time-dependent bed [9, 10, 38, 40, 41]. In present paper we propose to use the Riemann-solver which is adapted to the flow parameters for calculating shallow water flows over an arbitrary bed in the presence of external force. The proposed method belongs to the family of methods based on the solution of the dam-break problem. The method consists in reducing of the problem to successive solutions of classical shallow water equations on the flat plane using Godunov method with allowance for the vertical nonhomogeneity effect in calculating the fluxes through the boundaries of cells adjoining to stepwise boundaries. The vertical nonhomogeneity leads to the Riemann problem solution on a step based on the quasi-two-layer shallow water model developed in [29, 30, 31, 32, 47, 48]. We are solving the shallow-water equations for one layer, introducing the fictitious lower layer only as an auxiliary structure in setting up the appropriate Riemann problems for the upper layer. Besides quasi-two-layer approach leads to appearance of additional terms in one-layer finite-difference representation of balance equations. These terms provide the mechanical work made by nonhomogeneous bed interacting with flow. Quasi-two-layer model 29, 30, 31, 32, 47, 48 is extended here for a generalized time-dependent bed which represents an external force. The term time-dependent bed means that at each time moment bed can have different values.

The main difficulty in modeling of fluid flows over a complex bed consists in that both partly and complete flooded domains may take place. Partly flooded domains may appear in flows when fluid depth is related to the value of bed gradient when approximated by steps. In this case the step wall is partly wetting and the algorithm suggested in present work considers exactly mechanical work done only by this part of the step. Mechanical work done by nonhomogeneous bed means that the work is done in the process of interaction of non-stationary flow with nonhomogeneous bed, with the step and with part of the step. Partly flooded domains present a real challenge for most finite-difference schemes and require special efforts need to be made to capture such domains. Algorithm suggested in our work avoids this difficulty in a natural way. In present work we extend algorithms developed in [29, 30, 31, 32, 47, 48] to the case of multiply-connected domains including partly flooded and dry regions.

Section 2 presents the shallow water model over an arbitrary bed profile in the presence of external force. The corresponding shallow water equations are written and the idea of the suggested Riemann-solver adaptable to flow parameters is described. Section 3 presents the quasi-two-layer model and the developed finite-difference scheme. Section 4 reviews the well known finite-difference schemes for calculating shallow water flows accounting for the external force effect in the form of fictitious bed, and compares them with our scheme. We have demonstrated results of the computations that show the rate of formation of an adequate flux of the mass and the impulse. We compare our results with those obtained by the well-balancing method. Section 5 presents the results of test computations with a constant external force corresponding to an inclined bed. The simulation of large-scale motion of heavy fluid (gas) is performed with allowance for the Coriolis force over a complex trigonometric function shape bed. Calculations and comparisons with laboratory experiment of two-dimensional dam-break problem on a slope are performed. Interaction of the Tsunami wave with the shore line including an obstacle has been simulated. The main results of the work are formulated in section 6.

**2. Model of shallow water flows over an arbitrary bed in the presence of external force**

We consider two-dimensional shallow water equations for fluid flows over an arbitrary nonhomogeneous bed with the external force in the divergence form [49, 53]:

$$\begin{cases} \dfrac{\partial h}{\partial t} + \dfrac{\partial (hu)}{\partial x} + \dfrac{\partial (hv)}{\partial y} = 0 \\ \dfrac{\partial (hu)}{\partial t} + \dfrac{\partial \left(hu^2 + gh^2/2\right)}{\partial x} + \dfrac{\partial huv}{\partial y} = -gh\dfrac{db}{dx} + E_x h, \\ \dfrac{\partial (hv)}{\partial t} + \dfrac{\partial \left(hv^2 + gh^2/2\right)}{\partial y} + \dfrac{\partial hvu}{\partial x} = -gh\dfrac{db}{dy} + E_y h \end{cases} \quad (1)$$

where $h(x,y,t)$ is the depth of fluid, $u(x,y,t)$ is the depth-averaged horizontal velocity in the x direction, $v(x,y,t)$ is the depth-averaged horizontal velocity in the y direction, $b(x,y)$ is the function describing the bed shape, $\overline{E}(h,u,v,x,y,t) = (E_x, E_y)$ is the external force, $g$ is the gravity acceleration. The first equation in system (1) represents the mass conservation law; the second and third equations represent the momentum conservation laws for corresponding velocity vector components.

Let's introduce fluxes

$$F(\Psi) = \begin{pmatrix} hu \\ hu^2 + gh^2/2 \\ hvu \end{pmatrix}, \quad G(\Psi) = \begin{pmatrix} hv \\ huv \\ hv^2 + gh^2/2 \end{pmatrix},$$

where $\Psi = \begin{pmatrix} h \\ hu \\ hv \end{pmatrix}$,

and the source term $S = \begin{pmatrix} 0 \\ -gh\dfrac{db}{dx} + E_x h \\ -gh\dfrac{db}{dy} + E_y h \end{pmatrix}.$

Thus we rewrite equations (1) as:

$$\partial_t \Psi + \partial_x F(\Psi) + \partial_y G(\Psi) = S.$$

The corresponding one-dimensional problem is obtained by reducing the system (1) over one of spatial coordinates with an accuracy of projection of the momentum equation on a non-reduced spatial direction. This corresponds to the one-dimensional spatial problem of SWE over a time-dependent bed. Neglecting in (1), for certainty, the derivatives with respect to y we get the following equations:

$$\begin{cases} \dfrac{\partial h}{\partial t} + \dfrac{\partial (hu)}{\partial x} = 0 \\ \dfrac{\partial (hu)}{\partial t} + \dfrac{\partial \left(hu^2 + gh^2/2\right)}{\partial x} = -gh\dfrac{db}{dx} + E_x h \end{cases} \quad (2)$$

Here $u = u(x,t)$, $b = b(x)$ and $E_x = E_x(h,u,x,t)$. The problem becomes essentially two-dimensional if the projection of external force $E$ onto one of the coordinate axes depends either on the full velocity vector, or on the projection of a velocity vector onto other coordinate axis. Neglecting one of the spatial variables during the solving an essentially two-dimensional problem results in violating the momentum conservation law. This, in turn, makes necessary to introduce some fictitious work to compensate these violations, in spite of all non-physical character of such compensation. Such situation exists, for example in modeling of rotating shallow water, when Coriolis force acts as an external force. Use of quasi-two-layer model allows improving representation of transversal component of a velocity vector under the assumption of its stationarity on each time step that in particular leads to improvement of conservation of the potential vorticity vector. We will use this advantage of developed scheme in future developments. If *E* depends only on *h, u, x* and *t*, splitting of system (1) into a set of subsystems (2) is physically justified.

Existence of well-developed and validated numerical methods makes especially attractable the reduction of a problem in the presence an external force to solution of the problem of shallow water flows over a composed time-dependent bed. Actually, to obtain a formal coincidence of the types of equations, external force must be presented in the following form:

$$g\frac{dk}{dx} = -E_x, \quad (3)$$

Where $k = k(x,t)$ is the fictitious bed. As it was mentioned earlier $E_x$ depends on *h, u, x* and *t*. Function *k* depends only on *x* and *t* because *h* and *u* also depend only on *x* and *t*. Thus system (2) can be rewritten as:

$$\begin{cases} \dfrac{\partial h}{\partial t} + \dfrac{\partial (hu)}{\partial x} = 0 \\ \dfrac{\partial (hu)}{\partial t} + \dfrac{\partial \left(hu^2 + gh^2/2\right)}{\partial x} = -gh\dfrac{d(b+k)}{dx} \end{cases}. \quad (4)$$

Such a presentation of external force in shallow water equations essentially narrows the class of possible solutions, excluding from consideration spatially non-integrable functions specifying the external force $E$. This condition is generally needed to write down formally system (4), and may not be significant in finite difference representations of all functions defined in discrete points.

If we introduce the difference grid with a spatial mesh size $X$ in the $x$ direction and consider two adjacent cells in $x$, then, according to formula (3) we have in the left cell after averaging the fictitious bed with a slope tangent $E_l/g$, and with the tangent $E_r/g$ in the right cell. Therefore, the corresponding difference of average heights of a fictitious bed on the boundary of the two adjacent cells is:

$$\frac{|E_r X/g + E_l X/g|}{2} \qquad (5)$$

The possibility of simulating the shallow water in the presence of external force by introducing the fictitious time-dependent bed has been used for a long time for constructing numerical models [9, 10, 38, 40, 41]. The fact that the methods developed for composed bed can be useful in solving the problems with an external force, apparently, causes no doubts. In spite of the fact that the influence of fictitious bed, obviously, directly leads to accounting for bed work performed over the fluid flow, the external force, by itself, could not commit any work. The method proposed in this paper makes it possible to provide the physical interpretation of the mentioned formal adaptation of the methods. This makes it possible to better understand the limits of applicability of the given approach for splitting finite-difference schemes, being not limited to a particular applied model. The method proposed in this work allows to visualize the features of the splitting approach to numerical simulation as a whole and, thus, provides physical ideas for improving stability criteria in the finite-difference implementation.

The underlying problem for all Godunov-type methods is the Riemann problem that is the one-dimensional formulation of the Cauchy problem for two semi-infinite domains, in each of which the values of all hydrodynamic parameters are constant at the initial time. It is clear that in the presence of the external $y$-dependent force such a problem is physically meaningful for a rather small time intervals only, since its general solution is sought as a set of partial self-similar solutions.

As a result, the problem of calculating of the shallow water flows in the presence of external force becomes identical to the problem of calculating the shallow water flows over a time-dependent nonhomogeneous bed. Papers [29, 30, 31, 32, 47, 48] have offered the quasi-two-layer method for calculating shallow water flows over an nonhomogeneous bed. In present work we show how quasi-two-layer method can be applied in case of a time-dependent bed. The effective bed profile is approximated by a piecewise constant function separating it into a finite number of domains with a stepwise boundary.

## 3. Quasi-two-layer model and finite-difference scheme for the equations of shallow water over an arbitrary bed profile in the presence of external force

In this section we briefly describe basics of quasi-two-layer model [29, 31, 47] are required for understanding of suggested computational algorithm andr derive finite-difference scheme for the equations of shallow water over an arbitrary bed profile in the presence of external force. This model is extended here for non-stationary boundaries and for flows that include partly flooded and dry regions on complex bed. Distinctive feature of the suggested model is a separation of a studied flow into two layers in calculating flow quantities near each step, and improving by this means approximation of depth-averaged solutions of the initial three-dimensional Euler equations. The uniqueness of such a separation into two layers is provided by the uniqueness of solution of the Dirichlet problem for defining this boundary. One of the advantages of the developed method is that it allows one to take into account both flow velocity and the height of the step at each point of space and at any time instant. We are solving the shallow-water equations for one layer, introducing the fictitious lower layer only as an auxiliary structure in setting up the appropriate Riemann problems for the upper layer. Besides quasi-two-layer approach leads to appearance of additional terms in one-layer finite-difference representation of balance equations. These terms provide the mechanical work made by nonhomogeneous bed interacting with flow. Suggested approach modifies properly values of hydrodynamic fluxes in one-layer finite-difference scheme for nonhomogeneous bed as well by adjusting flow to that for upper layer in really two-layer model in the vicinity of discrete steps. Below we develop finite-difference representation of the external force in numerical Godunov-type models for flows of shallow water based on quasi-two-layer model [29, 31, 47].

### 3.1 Quasi-two-layer model

The basic idea of the quasi-two-layer model and model application for development of finite-difference scheme for shallow water equations over fully flooded bed is described in [30, 32, 48]. In this section we extend these ideas to the case of existence partially inundated areas in bed shape and to time dependent bed to take into account the external force influence on the flow.

We consider a shallow non-viscous flow over a step of height $b_0$ turned to the left with no loss of generality (Fig. 1). On fig.1 dashed line represents interface between layers.

If the fluid height is lower than the step height, the step would act as an impermeable boundary for the lower layer. Thus, if the height of a fluid is higher than that of the step, we may consider two fluid layers: the lower layer with the flow parameters $h_1$ and $u_1$ for which the step is an impermeable boundary and the upper one with the flow parameters $h_2$ and $u_2$ for which there is no direct influence of the step, obviously, $h_l = h_1 + h_2$. Equations for the two-layer system are following:

$$\begin{cases} \dfrac{\partial h_1}{\partial t} + \dfrac{\partial}{\partial x}(h_1 u_1) = 0, \\ \dfrac{\partial h_2}{\partial t} + \dfrac{\partial}{\partial x}(h_2 u_2) = 0, \\ \dfrac{\partial}{\partial t}(h_1 u_1) + \dfrac{\partial}{\partial x}(h_1 u_1^2 + \dfrac{1}{2} g h_1^2) + g h_1 \dfrac{\partial h_2}{\partial x} = 0, \\ \dfrac{\partial}{\partial t}(h_2 u_2) + \dfrac{\partial}{\partial x}(h_2 u_2^2 + \dfrac{1}{2} g h_2^2) + g h_2 \dfrac{\partial h_1}{\partial x} = 0, \end{cases} \quad (6)$$

with the corresponding boundary condition $u_1 = 0$ for $x = 0$, $t \geq 0$ (the step affects the height and velocity of the lower layer near the step) and with the initial conditions for $t = 0$:

$$h_1 = h^*, \ h_2 = h_l - h^*, \ u_1 = u_l, \ u_2 = u_l \ \text{for} \ x \leq 0,$$

$$h_2 = h_r, \ u_2 = u_r \qquad \text{for } x > 0. \tag{7}$$

To solve system (6) we split it into two so that the variables $h_1$ and $u_1$ can be calculated independently of the variables $h_2$ and $u_2$. Assuming that the wave pattern in the lower layer is formed much faster than that in the upper one, the layers are separated so that the lower layer interacting with the step is at rest and forms a common horizontal plane with the step. This makes it possible to disregard the term $gh_1 \partial h_2/\partial x$ in the third equation of system (6). We suppose that the upper layer does not appreciably affect the lower one, i.e., we disregard the term $gh_2 \partial h_1/\partial x$. Then the effect of the lower layer on the upper one is taken into account by changing the initial condition $h_2 = h_l - h^*, x \leq 0$, which is determined by the change in the lower layer depth due to the step drag. As confirmed below by numerical calculations, these assumptions allow us to describe adequately the resulting flow. The adequacy in this case should imply the compatibility between the numerical results and theoretically possible ones when analytically solving the problem of arbitrary discontinuity decay over the step for shallow water equations.

Thus, under the above assumptions, we have the quasi-two-layer model described by the equations where the variables $h_1, u_1$ and $h_2, u_2$ are now connected only by the initial parameter $h^*$ which should be chosen so that the lower layer height directly at the step coincides with the step height $h_1|_{x=0} = b_0$ for $t \geq 0$. Consequently, the problem of finding $h^*$ reduces to the solution of the Dirichlet inverse problem:

$$\begin{cases} \dfrac{\partial h_1}{\partial t} + \dfrac{\partial}{\partial x}(h_1 u_1) = 0, \\ \dfrac{\partial}{\partial t}(h_1 u_1) + \dfrac{\partial}{\partial x}(h_1 u_1^2 + \dfrac{1}{2}gh_1^2) = 0, \end{cases} \tag{8}$$

$h_1 = h^*$ for $x < 0$, $t = 0$; $u_1 = u_l$ for $x \leq 0$, $t = 0$;
$u_1 = 0$ for $x = 0$, $t \geq 0$; $h_1 = b_0$ for $x = 0$, $t \geq 0$.

The solution of system (8) depends on the fluid flow direction, that is, on velocity $u_l$:

a) $u_l > 0$, the flow has a form of the shock wave reflected to the left, on the left side of which the flow parameters are $h^*$ and $u_l > 0$. On the right side the fluid is at rest, that is, $h = b_0$ and $u = 0$, and the closing depth, determined by the step effect, is calculated as follows:

$$u_l = (b_0 - h^*)\sqrt{\dfrac{g}{2}\dfrac{(b_0 + h^*)}{b_0 h^*}}; \tag{9}$$

The solution is determined directly from the algebraic formula (9) that follows from Hugoniot's relations on a hydraulic jump [37];

b) $u_l < 0$, the flow has a form of rarefaction wave moving to the left, on the left side of which the flow parameters are $h^*$ and $u_l < 0$, and on the right side as $h = b_0$ and $u = 0$, Therefore, the solution is determined from the corresponding Riemann's invariant, which describes the rarefaction wave:

$$h^* = \frac{1}{g}\left(\sqrt{gb_0} - \frac{1}{2}u_l\right)^2. \qquad (10)$$

Expression (10) determines, in the explicit form, the depth of the lower part of a flow that is completely stopped by the step.

Thus, we have found $h^*$ and for $x \leq 0$ we now have: $h_1 = h^*$ and $h_2 = h_l - h^*$. The assumption that the wave pattern in the lower layer is formed much faster than that in the upper one is not valid for the case of a rather small step compared to the other flow parameters. In this situation it is necessary to assume that the wave pattern in the upper layer is formed faster than that in the lower one. The relation between the layers' heights has the form: $h_2 = h^*$ and $h_1 = h_l - h^*$. For the boundary situation between the above possibilities, correct splitting of the system becomes impossible, the two-layer model no longer adequately describes the physical process and it is necessary to use the complete Euler equations as a minimum for the fluid layer in the $\varepsilon$-neighbourhood of the step boundary edge.

The situation is essentially different in the case when $h^*$ is outside interval $(0, h_l]$. Physically, it is possible in two situations. First, $h^*$ may be equal to zero. This implies absence of the lower layer, and consequently absence of the fluid and/or changes of the height of the bed. Second possibility is $h^* > h_l$. It corresponds to the full deceleration of the layer at the left side, i.e. in such situation the bed is a non-leaking boundary for all fluid at the left and therefore, it is necessary to accept that $h_2 = 0$ for $x < 0$. Then value $h_1$ for $x = 0$ can be found as a solution of the problem of interaction of all fluid at the left with a non-leaking boundary.

After evaluation of $h^*$ one can find the values of flow quantities as the solution of classical Riemann problem for shallow water over a smooth bed with corresponding initial conditions. Using the quasi-two-layer method, one can find the flow quantities on cell faces, which determine, in the difference scheme, the values of all hydrodynamic parameters in the next time layer.

Thus, depending on flow parameters near the stepwise bed, the depths of fluid's lower and upper layers are calculated, and for calculating the flow parameters on a face the classical Riemann problem for upper layer's parameters is solved. The proposed method is adapted to the flow parameters and allows one to take into account the fluid flow features at each point of space and at each time instant. It should be noted that these features are considered in suggested model only in two-layer approximation in the steps vicinity.

### *3.2. Derivation of finite difference scheme*

To obtain the difference scheme we integrate the system of equations of two-layer shallow water with the external force over a nonhomogeneous bed (1):

$$\begin{cases} \iiint\limits_G \left( \frac{\partial(h_1+h_2)}{\partial t} + \frac{\partial(h_1 u_1)}{\partial x} + \frac{\partial(h_2 u_2)}{\partial x} + \frac{\partial(h_1 v_1)}{\partial y} + \frac{\partial(h_2 v_2)}{\partial y} \right) dxdydt = 0 \\ \iiint\limits_G \left( \begin{array}{l} \frac{\partial(h_1 u_1)}{\partial t} + \frac{\partial(h_1 u_1^2 + gh_1^2/2)}{\partial x} + \frac{\partial h_1 u_1 v_1}{\partial y} + \frac{\partial(h_2 u_2)}{\partial t} + \frac{\partial(h_2 u_2^2 + gh_2^2/2)}{\partial x} + \\ + \frac{\partial h_2 u_2 v_2}{\partial y} + g(h_1+h_2)\frac{db_x}{dx} - E_x(h_1+h_2) + g\frac{\partial h_2 h_1}{\partial x} \end{array} \right) dxdydt = 0 \\ \iiint\limits_G \left( \begin{array}{l} \frac{\partial(h_1 v_1)}{\partial t} + \frac{\partial(h_1 v_1^2 + gh_1^2/2)}{\partial y} + \frac{\partial h_1 v_1 u_1}{\partial x} + \frac{\partial(h_2 v_2)}{\partial t} + \frac{\partial(h_2 v_2^2 + gh_2^2/2)}{\partial y} + \\ + \frac{\partial h_2 v_2 u_2}{\partial x} + g(h_1+h_2)\frac{db_y}{dy} - E_y(h_1+h_2) + g\frac{\partial h_2 h_1}{\partial y} \end{array} \right) dxdydt = 0 \end{cases}, \quad (11)$$

where $G(x,y,t)$ is the arbitrary nonempty domain, which is homeomorphic to a spatial cube, $h_1, u_1, v_1$ are lower layer parameters, $h_2, u_2, v_2$ are upper layer parameters. Passing in (11) to surface integrals, we obtain the following integral form:

$$\begin{cases} \oiint\limits_S (h_1+h_2) dxdy + \oiint\limits_S h_1 u_1 dydt + \oiint\limits_S h_1 v_1 dxdt + \oiint\limits_S h_2 u_2 dydt + \oiint\limits_S h_2 v_2 dxdt = 0 \\ \oiint\limits_S (h_1 u_1 + h_2 u_2) dxdy + \oiint\limits_S (h_1 u_1^2 + gh_1^2/2 + h_2 u_2^2 + gh_2^2/2 + gh_1 h_2) dydt + \\ + \oiint\limits_S (h_1 u_1 v_1 + h_2 u_2 v_2) dxdt - \iiint\limits_G \left( E_x(h_1+h_2) - g(h_1+h_2)\frac{db_x}{dx} \right) dxdydt = 0 \\ \oiint\limits_S (h_1 v_1 + h_2 v_2) dxdy + \oiint\limits_S (h_1 v_1^2 + gh_1^2/2 + h_2 v_2^2 + gh_2^2/2 + gh_1 h_2) dxdt + \\ + \oiint\limits_S (h_1 v_1 u_1 + h_2 v_2 u_2) dydt - \iiint\limits_G \left( E_y(h_1+h_2) - g(h_1+h_2)\frac{db_y}{dy} \right) dxdydt = 0 \end{cases}, \quad (12)$$

where $S$ is the boundary of domain $G(x,y,t)$. We apply the integral conservation laws (12) to each cell, choosing $S$ to be the surface determined by cell boundary at the time step $\tau$. Taking into account the evolution of the lower layer one gets the following equations on the cell faces:

$$x = x - 1/2 \Rightarrow h_2 = \tilde{H}, h_1 = k_{x-1/2} + b_{x-1/2}, u_1 = 0, u_2 = \tilde{U};$$
$$x = x + 1/2 \Rightarrow h_2 = \tilde{H}, h_1 = k_{x+1/2} + b_{x+1/2}, u_1 = 0, u_2 = \tilde{U};$$
$$y = y - 1/2 \Rightarrow h_2 = \tilde{H}, h_1 = k_{y-1/2} + b_{y-1/2}, v_1 = 0, v_2 = \tilde{V};$$
$$y = y + 1/2 \Rightarrow h_2 = \tilde{H}, h_1 = k_{y+1/2} + b_{y+1/2}, v_1 = 0, v_2 = \tilde{V};$$
$$\text{and} \quad h_1 + h_2 = \tilde{H} \quad \text{inside the cell.}$$

Here $x \pm 1/2, y \pm 1/2$ denote the coordinates for the boundary of the cell with a number (x,y). Here $k_{x-1/2} = k_x - k_{x-1}$, $b_{x-1/2} = b_x - b_{x-1}$, $k_{x+1/2} = k_{x+1} - k_x$, $b_{x+1/2} = b_{x+1} - b_x$, $k_{y-1/2} = k_y - k_{y-1}$, $b_{y-1/2} = b_y - b_{y-1}$, $k_{y+1/2} = k_{y+1} - k_y$, $b_{y+1/2} = b_{y+1} - b_y$. Besides, we apply the identity $gh_1^2/2 + gh_2^2/2 + gh_1 h_2 \equiv g(h_1 + h_2)^2/2$. Taking into account that fact that the bed and external force effects are now described by the height of the lower layer $h_1$ (functions b(x,y) and k(x,y,t) correspondingly) one gets

$$\begin{cases} \int_{x-1/2}^{x+1/2} \int_{y-1/2}^{y+1/2} \tilde{H}\big|_{t=t+\tau} dxdy - \int_{x-1/2}^{x+1/2} \int_{y-1/2}^{y+1/2} \tilde{H}\big|_{t=t} dxdy + \int_{y-1/2}^{y+1/2} \int_{t}^{t+\tau} \tilde{H}\tilde{U}\big|_{x=x+1/2} dydt - \int_{y-1/2}^{y+1/2} \int_{t}^{t+\tau} \tilde{H}\tilde{U}\big|_{x=x-1/2} dydt + \\ + \int_{x-1/2}^{x+1/2} \int_{t}^{t+\tau} \tilde{H}\tilde{V}\big|_{y=y+1/2} dxdt - \int_{x-1/2}^{x+1/2} \int_{t}^{t+\tau} \tilde{H}\tilde{U}\big|_{y=y-1/2} dxdt = 0 \\ \int_{x-1/2}^{x+1/2} \int_{y-1/2}^{y+1/2} \tilde{H}\tilde{U}\big|_{t=t+\tau} dxdy - \int_{x-1/2}^{x+1/2} \int_{y-1/2}^{y+1/2} \tilde{H}\tilde{U}\big|_{t=t} dxdy + \int_{y-1/2}^{y+1/2} \int_{t}^{t+\tau} \left( h_2 \tilde{U}^2 + \frac{1}{2} g \left( h_2 + k_{x+1/2} + b_{x+1/2} \right)^2 \right)\bigg|_{x=x+1/2} dydt - \\ - \int_{y-1/2}^{y+1/2} \int_{t}^{t+\tau} \left( h_2 \tilde{U}^2 + \frac{1}{2} g \left( h_2 + k_{x-1/2} + b_{x-1/2} \right)^2 \right)\bigg|_{x=x-1/2} dydt + \int_{x-1/2}^{x+1/2} \int_{t}^{t+\tau} h_2 \tilde{U}\tilde{V}\big|_{y=y+1/2} dxdt - \int_{x-1/2}^{x+1/2} \int_{t}^{t+\tau} \tilde{H}\tilde{U}\tilde{V}\big|_{y=y-1/2} dxdt = 0 \\ \int_{x-1/2}^{x+1/2} \int_{y-1/2}^{y+1/2} \tilde{H}\tilde{V}\big|_{t=t+\tau} dxdy - \int_{x-1/2}^{x+1/2} \int_{y-1/2}^{y+1/2} \tilde{H}\tilde{V}\big|_{t=t} dxdy + \int_{x-1/2}^{x+1/2} \int_{t}^{t+\tau} \left( h_2 \tilde{V}^2 + \frac{1}{2} g \left( h_2 + k_{y+1/2} + b_{y+1/2} \right)^2 \right)\bigg|_{y=y+1/2} dxdt - \\ - \int_{x-1/2}^{x+1/2} \int_{t}^{t+\tau} \left( h_2 \tilde{V}^2 + \frac{1}{2} g \left( h_2 + k_{y-1/2} + b_{y-1/2} \right)^2 \right)\bigg|_{y=y-1/2} dxdt + \int_{y-1/2}^{y+1/2} \int_{t}^{t+\tau} h_2 \tilde{V}\tilde{U}\big|_{x=x+1/2} dydt - \int_{y-1/2}^{y+1/2} \int_{t}^{t+\tau} \tilde{H}\tilde{V}\tilde{U}\big|_{x=x-1/2} dydt = 0 \end{cases}$$

(13)

Applying in (13) the mean value theorem and supposing that at cell boundary the values of all hydrodynamic parameters represent the solutions of a corresponding one-dimensional problem during the whole integration time step of, we obtain the difference scheme:

$$\frac{H_{x,y}^{t+\tau} - H_{x,y}^{t}}{\tau} = \frac{H_{x-1/2,y}^{t} U_{x-1/2,y}^{t} - H_{x+1/2,y}^{t} U_{x+1/2,y}^{t}}{X} + \frac{H_{x,y-1/2}^{t} V_{x,y-1/2}^{t} - H_{x,y+1/2}^{t} V_{x,y+1/2}^{t}}{Y},$$

$$\frac{H_{x,y}^{t+\tau} U_{x,y}^{t+\tau} - H_{x,y}^{t} U_{x,y}^{t}}{\tau} = \left( \begin{array}{c} \dfrac{g\left( H_{x-1/2,y}^{t} + i_x \times \left( \left( B_{x,y}^{t} - B_{x-1,y}^{t} \right) + \left( K_{x,y}^{t} + K_{x-1,y}^{t} \right)/2 \right) \right)^2}{2} - \\ -\dfrac{g\left( H_{x+1/2,y}^{t} + i_x \times \left( \left( B_{x+1,y}^{t} - B_{x,y}^{t} \right) + \left( K_{x+1,y}^{t} + K_{x,y}^{t} \right)/2 \right) \right)^2}{2} \\ + H_{x-1/2,y}^{t} \left( U_{x-1/2,y}^{t} \right)^2 - H_{x+1/2,y}^{t} \left( U_{x+1/2,y}^{t} \right)^2 \end{array} \right) / X +$$

$$+ \left( H_{x,y-1/2}^{t} U_{x,y-1/2}^{t} V_{x,y-1/2}^{t} - H_{x,y+1/2}^{t} U_{x,y+1/2}^{t} V_{x,y+1/2}^{t} \right)/Y$$

(14)
$$\frac{H^{t+\tau}_{x,y}V^{t+\tau}_{x,y} - H^{t}_{x,y}V^{t}_{x,y}}{\tau} = \left\{\begin{array}{c} \dfrac{g\left(H^{t}_{x,y-1/2} + i_y \times \left(\left(B^{t}_{x,y} - B^{t}_{x,y-1}\right) + \left(K^{t}_{x,y} + K^{t}_{x,y-1}\right)/2\right)\right)^2}{2} - \\ -\dfrac{g\left(H^{t}_{x,y+1/2} + i_y \times \left(\left(B^{t}_{x,y+1} - B^{t}_{x,y}\right) + \left(K^{t}_{x,y+1} + K^{t}_{x,y}\right)/2\right)\right)^2}{2} + \\ + H^{t}_{x,y-1/2}\left(U^{t}_{x,y-1/2}\right)^2 - H^{t}_{x,y+1/2}\left(U^{t}_{x,y+1/2}\right)^2 \end{array}\right\}/Y +$$

$$+ \left(H^{t}_{x-1/2,y}U^{t}_{x-1/2,y}V^{t}_{x-1/2,y} - H^{t}_{x+1/2,y}U^{t}_{x+1/2,y}V^{t}_{x+1/2,y}\right)/X$$

Here $\tau$ is the time step; $X$ and $Y$ are spatial mesh sizes; $H$ is the depth of fluid; $U$ is the velocity in the $x$ direction; $V$ is the velocity in the $y$ direction. $H^{t}_{x,y}, U^{t}_{x,y}, V^{t}_{x,y}$ are the space averages of $\tilde{H}, \tilde{U}, \tilde{V}$ respectively. $H^{t}_{x\pm1/2,y\pm1/2}, U^{t}_{x\pm1/2,y\pm1/2}, V^{t}_{x\pm1,y\pm1}$ are the time averages of $\tilde{H}, \tilde{U}, \tilde{V}$ respectively. Subscripts $x, y$ designate the values of function related to the center of masses of a cell with number $(x, y)$. Half-integers $x\pm1/2, y\pm1/2$ designate the values of quantities at the boundary between cells with numbers $x$, $x\pm1$, and $y$, $y\pm1$, respectively. Superscript $t$ designates the number of a step in time, $K_{x,y}$ is the height of a fictitious bed and $B_{x,y}$ is the height of a bed. Respectively, according to formula (5):

$$\frac{K_{x+1,y} + K_{x,y}}{2} = \frac{\left|E_{x+1,y}X/g + E_{x,y}X/g\right|}{2}, \quad \frac{K_{x,y} + K_{x-1,y}}{2} = \frac{\left|E_{x,y}X/g + E_{x-1,y}X/g\right|}{2},$$

$$\frac{K_{x,y+1} + K_{x,y}}{2} = \frac{\left|E_{x,y+1}X/g + E_{x,y}X/g\right|}{2}, \quad \frac{K_{x,y} + K_{x,y-1}}{2} = \frac{\left|E_{x,y}X/g + E_{x,y-1}X/g\right|}{2}.$$

The variables $i_x, i_y$ assume either the value 0 – in the case of negative difference of corresponding heights $\left(B_{x+1,y} - B_{x,y}\right) + \left(K_{x+1,y} + K_{x,y}\right)/2$, $\left(B_{x,y+1} - B_{x,y}\right) + \left(K_{x,y+1} + K_{x,y}\right)/2$, $\left(B_{x,y} - B_{x-1,y}\right) + \left(K_{x,y} + K_{x-1,y}\right)/2$, $\left(B_{x,y} - B_{x,y-1}\right) + \left(K_{x,y} + K_{x,y-1}\right)/2$ of a effective bed (is equal sum of bed and fictitious bed), or the value of $0 \leq i_x, i_y \leq 1$ – in the case of positive difference. Variables $i_x$, $i_y$ takes value 1 in case if $h^*$ is determined by equations (9,10) for a corresponding face of a cell and does not exceed value of depth inside the cell. Otherwise, variables $i_x$, $i_y$ on a corresponding face are equal to the ratio of the depth formed under the total retardation of the flow on the indicated face to the corresponding difference of heights of the effective bed. Thus unlikely algorithms developed in [29,30,31,32,47,48], in finite-difference scheme (14) party flooded and dry steps of effective bed are accounted. The values $H^{t}_{x\pm1/2,y\pm1/2}$, $U^{t}_{x\pm1/2,y\pm1/2}$, $V^{t}_{x\pm1/2,y\pm1/2}$ are calculated on the faces by solving the corresponding Riemann problems considering that transversal flow velocity is transported convectively. We use convective transportation of the velocity hypothesis in 1D problem instead of using 3$^{rd}$ equation. It is clear that developed finite-difference scheme is well-balanced type since it is balanced in case of zero velocities and free surface is horizontal.

If we rewrite (14) in the terms of fluxes $F(\Psi)$ and $G(\Psi)$ we get:

$$\Psi_{xy}^{t+1} = \Psi_{xy}^{t} - \frac{\Delta t}{\Delta x}\left(F_{x+1/2,y} - F_{x-1/2,y} + FBK_{x+1/2,y} - FBK_{x-1/2,y}\right) -$$

$$-\frac{\Delta t}{\Delta y}\left(G_{x,y+1/2} - G_{x,y-1/2} + FBK_{x,y+1/2} - FBK_{x,y-1/2}\right),$$

where

$$FBK_{x+1/2,y} = \begin{pmatrix} 0 \\ g\left(H_{x+1/2,y}^{t} \times i_x \times \left(\left(B_{x+1,y}^{t} - B_{x,y}^{t}\right) + \left(K_{x+1,y}^{t} + K_{x,y}^{t}\right)/2\right)\right) + \\ +\frac{g\left(i_x \times \left(\left(B_{x+1,y}^{t} - B_{x,y}^{t}\right) + \left(K_{x+1,y}^{t} + K_{x,y}^{t}\right)/2\right)\right)^2}{2} \\ 0 \end{pmatrix},$$

$$FBK_{x-1/2,y} = \begin{pmatrix} 0 \\ g\left(H_{x-1/2,y}^{t} \times i_x \times \left(\left(B_{x,y}^{t} - B_{x-1,y}^{t}\right) + \left(K_{x,y}^{t} + K_{x-1,y}^{t}\right)/2\right)\right) + \\ +\frac{g\left(i_x \times \left(\left(B_{x,y}^{t} - B_{x-1,y}^{t}\right) + \left(K_{x,y}^{t} + K_{x-1,y}^{t}\right)/2\right)\right)^2}{2} \\ 0 \end{pmatrix},$$

$$FBK_{x,y+1/2} = \begin{pmatrix} 0 \\ 0 \\ g\left(H_{x,y+1/2}^{t} \times i_y \times \left(\left(B_{x,y+1}^{t} - B_{x,y}^{t}\right) + \left(K_{x,y+1}^{t} + K_{x,y}^{t}\right)/2\right)\right) + \\ +\frac{g\left(i_y \times \left(\left(B_{x,y+1}^{t} - B_{x,y}^{t}\right) + \left(K_{x,y+1}^{t} + K_{x,y}^{t}\right)/2\right)\right)^2}{2} \end{pmatrix},$$

$$FBK_{x,y-1/2} = \begin{pmatrix} 0 \\ 0 \\ g\left(H^t_{x,y-1/2} \times i_y \times \left(\left(B^t_{x,y} - B^t_{x,y-1}\right) + \left(K^t_{x,y} + K^t_{x,y-1}\right)/2\right)\right) + \\ +\dfrac{g\left(i_y \times \left(\left(B^t_{x,y} - B^t_{x,y-1}\right) + \left(K^t_{x,y} + K^t_{x,y-1}\right)/2\right)\right)^2}{2} \end{pmatrix}.$$

The studied equations are hyperbolic type. Therefore, by the analogy with the shallow water equations, in which the external force effects is disregarded [37], the developed method uses the standard Courant – Friedrichs condition as the stability condition. This condition, however, should take into consideration the velocity of propagation of perturbations, including those in the fictitious lower layer. That is, the time step cannot exceed the minimal time during which the perturbations in each layer pass the half of a cell. With regard to correction for two-dimensional statement we have:

$$\tau = R\frac{\Delta t_x \Delta t_y}{\Delta t_x + \Delta t_y}, \qquad (15)$$

where $\Delta t_x, \Delta t_y$ are the minimum times of propagation of perturbations along the abscissa and ordinate axes, respectively; $R < 1$ is the additional factor for increasing the reliability. In calculations presented below this factor is taken equal to 0.4. The success of selecting the stability criterion in this form is confirmed by test calculations presented below.

For improving accuracy of developed finite-difference scheme we use 2$^{nd}$ order accuracy computations in some tests (Subsections 5.2,5.3,5.4). Increase of accuracy on spatial coordinate is reached by applying the piecewise-linear reconstruction to the distribution of functions value in a cell with the use of the minmod limiter suggested for the first time by Kolgan [36,37] for the solutions of accuracy problems of Godunov type methods in gas-dynamics:

$$W^t_x = \min\mathrm{mod}\left(\frac{F^t_{x+1} - F^t_x}{\Delta x}, \frac{F^t_x - F^t_{x-1}}{\Delta x}\right) \qquad \text{where } F^t_x \equiv \begin{pmatrix} H^t_x \\ U^t_x \\ V^t_x \end{pmatrix}, \; \alpha = 0.72.$$

$$\min\mathrm{mod}(a,b) = \alpha\frac{1}{2}(sign\,a + sign\,b)\min(|a|,|b|)$$

The second order of accuracy in time is reached by application of two-step-by-step algorithm predictor - corrector. At a stage predictor there are found auxiliary values of required sizes for the whole step on time with the help of quasi-two-layer algorithm of the first order of the accuracy. These auxiliary values are used to find the values on an intermediate step in time by using arithmetic averaging with values of the previous time step. On the corrector step the given sizes are reconstructed in space: $F_x^{t+\frac{\tau}{2}} + \frac{1}{2}\Delta x W_x^t$, $F_{x+1}^{t+\frac{\tau}{2}} - \frac{1}{2}\Delta x W_{x+1}^t$, accordingly at the left and at the right sides of the face $x+1/2$. Next, the values of fluid variables on borders of cells corresponding to an intermediate time layer are found.

## 4. Analysis and comparison of the different methods for modeling of shallow water flows in presence the external force

### 4.1. Theoretical comparisions

In this section we compare our algorithm with the other finite-difference schemes for numerical simulations of shallow water flows with in the presence of external force in the form of a fictitious bed. To simplify the presentation, all derivations are performed in the cases of one-dimensional finite-difference schemes. We consider the case when the cell adjoins the step of height $b_0$ on the left side. We shall write down the finite-difference scheme for the wave-propagation method [40, 41] in the explicit form. The method is based on solution of the auxiliary Riemann problem at the center of each cell where the difference of fluxes is selected in such a way that the external force effect is exactly balanced in the stationary case. Thus, at the predictor step, the flow parameters $h_{x-}, u_{x-}$ and $h_{x+}, u_{x+}$ are calculated inside a cell. Then, at the corrector step, using the flow parameters obtained at the predictor step, the flow parameters on cell's interfaces $h_{i-1/2}, u_{i-1/2}$ and $h_{x+1/2}, u_{x+1/2}$ are calculated (Fig. 2). In fact, such an approach corresponds to the momentum and mass redistribution inside the considered cell, in order to compensate the influence of a step. The presence of such a sub grid discontinuity ensures the description of the non-divergent right-hand side of the system of equations (2) owing to the additional source of momentum. As a result, the standard system of shallow water equations over the smooth bed with modified fluxes through the interfaces of cells is considered.

To obtain the finite-difference scheme, we integrate over each of cell parts obtained by fictitious partition of the whole cell at the center:

$$\begin{cases} \oiint_{G_{left}} \left[ \frac{dh}{dt} + \frac{d(hu)}{dx} \right] dxdt + \oiint_{G_{right}} \left[ \frac{dh}{dt} + \frac{d(hu)}{dx} \right] dxdt = 0 \\ \oiint_{G_{left}} \left[ \frac{\partial}{\partial t}(hu) + \frac{\partial}{\partial x}(hu^2 + \frac{1}{2}gh^2) \right] dxdt + \oiint_{G_{right}} \left[ \frac{\partial}{\partial t}(hu) + \frac{\partial}{\partial x}(hu^2 + \frac{1}{2}gh^2) \right] dxdt = 0 \end{cases} \quad (16)$$

The application of the Ostrogradsky-Gauss formulas transforms (16) to the form:

$$\begin{cases} \int\limits_{x-1/2}^{x} h\big|_{t=t+1} dx + \int\limits_{x}^{x+1/2} h\big|_{t=t+1} dx - \int\limits_{x-1/2}^{x} h\big|_{t=t} dx - \int\limits_{x}^{x+1/2} h\big|_{t=t} dx + \\ + \int\limits_{t}^{t+1} (hu)\big|_{x+1/2} dt - \int\limits_{t}^{t+1} (hu)\big|_{x-1/2} dt + \int\limits_{t}^{t+1} (hu)\big|_{x+} dt - \int\limits_{t}^{t+1} (hu)\big|_{x-} dt = 0 \\ \int\limits_{x-1/2}^{x} (hu)\big|_{t=t+1} dx + \int\limits_{x}^{x+1/2} (hu)\big|_{t=t+1} dx - \int\limits_{x-1/2}^{x} (hu)\big|_{t=t} dx - \int\limits_{x}^{x+1/2} (hu)\big|_{t=t} dx + \\ + \int\limits_{t}^{t+1} \left(hu^2 + \frac{1}{2}gh^2\right)\bigg|_{x+1/2} dt - \int\limits_{t}^{t+1} \left(hu^2 + \frac{1}{2}gh^2\right)\bigg|_{x-1/2} dt + \\ + \int\limits_{t}^{t+1} \left(hu^2 + \frac{1}{2}gh^2\right)\bigg|_{x-} dt - \int\limits_{t}^{t+1} \left(hu^2 + \frac{1}{2}gh^2\right)\bigg|_{x+} dt = 0 \end{cases} \quad (17)$$

The use of the mean value theorem and performance of integration allows one to rewrite system (17) in the form:

$$\begin{cases} H_x^{t+1}\frac{1}{2}\Delta x + H_x^{t+1}\frac{1}{2}\Delta x - H_x^{t}\frac{1}{2}\Delta x - H_x^{t}\frac{1}{2}\Delta x + H_{x+1/2}^{t}U_{x+1/2}^{t}\tau - H_{x-1/2}^{t}U_{x-1/2}^{t}\tau = 0 \\ H_x^{t+1}U_x^{t+1}\frac{1}{2}\Delta x + H_x^{t+1}U_x^{t+1}\frac{1}{2}\Delta x - H_x^{t}U_x^{t}\frac{1}{2}\Delta x - H_x^{t}U_x^{t}\frac{1}{2}\Delta x + \\ + (H_{x+1/2}^{t}U_{x+1/2}^{t}{}^2 + \frac{1}{2}gH_{x+1/2}^{t}{}^2)\tau - (H_{x-1/2}^{t}U_{x-1/2}^{t}{}^2 + \frac{1}{2}gH_{x-1/2}^{t}{}^2)\tau + \\ + \left(\left(H_x^{t} + \frac{b_0}{2}\right)U_{x+}^{t}{}^2 + \frac{1}{2}g\left(H_x^{t} + \frac{b_0}{2}\right)^2\right)\tau - \left(\left(H_x^{t} - \frac{b_0}{2}\right)U_{x+}^{t}{}^2 + \frac{1}{2}g\left(H_x^{t} - \frac{b_0}{2}\right)^2\right)\tau = 0 \end{cases} \quad (18)$$

Taking into account that in the cell center the stationary problem is solved, such that $H_{x+}^{t} = H_x^{t} + \frac{1}{2}b_0$, $H_{x-}^{t} = H_x^{t} - \frac{1}{2}b_0$, and in the small neighborhood $\varepsilon$: $U_{x+} = U_{x-} = 0$ [40,41], system (18) is transformed into the following finite-difference scheme:

$$\begin{cases} H_x^{t+1} = H_x^{t} + \tau \times \left(\dfrac{H_{x-1/2}^{t}U_{x-1/2}^{t} - H_{x+1/2}^{t}U_{x+1/2}^{t}}{X}\right) \\ U_x^{t+1} = \tau \times \begin{pmatrix} H_{x-1/2}^{t}U_{x-1/2}^{t}{}^2 + \dfrac{1}{2}gH_{x-1/2}^{t}{}^2 - \\ -H_{x+1/2}^{t}U_{x+1/2}^{t}{}^2 - \dfrac{1}{2}gH_{x+1/2}^{t}{}^2 - \\ -gH_x^{t}b_0 \end{pmatrix} / XH_x^{t+1} + \dfrac{H_x^{t}U_x^{t}}{H_x^{t+1}} \end{cases} \quad (19)$$

In particular case when one-dimensional problem is considered and step height is $b_0$, the second equation of the obtained finite-difference scheme formally coincides with the second equation of the quasi-two-layer finite-difference scheme (14) under the following condition:

$$H_x = H_{x+1/2} + \frac{b_0}{2} \quad (20)$$

where the flux quantity $H_{x+1/2}$ is calculated by the quasi-two-layer method.

Note that in the wave-propagation method the flow parameters at the cell boundary $H_{x+1/2}, U_{x+1/2}$ are calculated as the solution of the dam-break problem for the initial parameters on the left $H_x + \frac{b_0}{2}, U_x$ and on the right $H_{x+1}, U_{x+1}$ sides of the discontinuity plane. In the quasi-two-layer method $H_{x+1/2}$ is calculated as the solution of the dam-break problem for the initial parameters on the left $H_x - H^*, U_x$, and on the right $H_{x+1}, U_{x+1}$ sides of the discontinuity plane.

The finite-difference scheme proposed in papers [3,9,10,44] also interprets the term containing the external force as some fictitious bed. The method proposed in papers [3,9,10,44] is based on the scheme of hydrostatic reconstruction of the flows on cell interfaces, which uses the solutions of steady states. In the one-dimensional case the finite-difference scheme is obtained from the classical scheme for a smooth bed by adding the source terms $\frac{1}{2}gH^t_{x+1/2-}{}^2, \frac{1}{2}gH^t_{x-1/2+}{}^2$ into the scheme. On the interfaces, however, the corrected flow parameters are used, which take into account the hydrostatic balance. The scheme looks as follows:

$$\begin{cases} H_x^{t+1} = H_x^t + \tau \times \left( \dfrac{H^t_{x-1/2}U^t_{x-1/2} - H^t_{x+1/2}U^t_{x+1/2}}{X} \right) \\ U_x^{t+1} = \tau \times \left( \begin{array}{l} H^t_{x-1/2}U^t_{x-1/2}{}^2 + \dfrac{1}{2}gH^t_{x-1/2}{}^2 - H^t_{x+1/2}U^t_{x+1/2}{}^2 - \\ -\dfrac{1}{2}gH^t_{x+1/2}{}^2 + \dfrac{1}{2}gH^t_{x+1/2-}{}^2 - \dfrac{1}{2}gH^t_{x-1/2+}{}^2 \end{array} \right) / XH_x^{t+1} + \dfrac{H_x^t U_x^t}{H_x^{t+1}} \end{cases} \quad (21)$$

Using the hydrostatic balance condition [3,9,10,44], we find that in the case of a step of height $b_0$, on the right side of the considered cell $H^t_{x+1/2-} = H^t_x - b_0$, $H^t_{x-1/2+} = H^t_x$. Then the finite-difference scheme will be re-written in the form of:

$$\begin{cases} H_x^{t+1} = H_x^t + \tau \times \left( \dfrac{H^t_{x-1/2}U^t_{x-1/2} - H^t_{x+1/2}U^t_{x+1/2}}{X} \right) \\ U_x^{t+1} = \tau \times \left( \begin{array}{l} H^t_{x-1/2}U^t_{x-1/2}{}^2 + \dfrac{1}{2}gH^t_{x-1/2}{}^2 - H^t_{x+1/2}U^t_{x+1/2}{}^2 - \\ -\dfrac{1}{2}gH^t_{x+1/2}{}^2 + \dfrac{1}{2}gb_0{}^2 - gH^t_x b_0 \end{array} \right) / XH_x^{t+1} + \dfrac{H_x^t U_x^t}{H_x^{t+1}} \end{cases} \quad (22)$$

The second equation of scheme (22), as well as that of scheme (19), formally coincides with the second equation of scheme (14), provided that:

$$H_x = H_{x+1/2} + b_0 \qquad (23)$$

where the flux value $H_{x+1/2}$ is calculated by the quasi-two-layer method.

Note that in the method proposed in papers [3,9,10,44] $H_{x+1/2}, U_{x+1/2}$ is calculated as the solution of the dam-break problem for the initial parameters on the left $H_x - b_0, U_x$, and on the right $H_{x+1}, U_{x+1}$ sides of the discontinuity plane, whereas in the quasi-two-layer method $H_{x+1/2}$ is calculated as the solution of the dam-break problem for the initial parameters on the left $H_x - H^*, U_x$, and on the right $H_{x+1}, U_{x+1}$ sides of the discontinuity plane.

Generally speaking, the finite-difference scheme proposed in papers [3, 9, 10, 44] can be obtained in the same manner, as the LeVeque scheme [40,41]. Indeed, it is sufficient to consider, at the predictor step in the LeVeque method [40,41], the problem with additional source corresponding to the well-balance principle as the stationary hydrostatic problem (Figure 2, wave-propagation algorithm [40,41] – the dotted line, and the well-balancing algorithm [3, 9, 10, 44] – the dash-dotted line). It is important to emphasize that in paper [5], devoted to numerical simulation of the classical Saint-Venant system, a similar approach of mass redistribution inside a cell gave quite satisfactory results. Unlike the situation with a non-differentiable nonhomogeneity of the right-hand side, in the given method the introduction of additional discontinuity decay inside a cell is not required. Definitely, after integration of corresponding differential equations, the obtained balance equations contain the necessary term, which describe the work made of the adjacent bed slope in the explicit form.

Taking into consideration that the finite-difference presentations for the momentum conservation law with regard to (20) and (23) formally coincide our quasi-two-layer algorithm is extension both methods up to the accuracy of the method of finding flow quantities on cell interfaces. In this model $H_{x+1/2}$ is calculated with using (for the considered problem) the flow depth $H_x - H^*$ on the left side from a discontinuity and $H_{x+1}$ – on the right side. Here $H^*$, depending on the flow conditions, is chosen as a function of current hydrodynamic values. One can assume, in conformity with the physical situation, the whole spectrum of values $[0, b_0]$ – at flow running on the fictitious bed, and $[b_0, H_x]$ – in the opposite case. At the same time, in the considered models the correction of flow quantities on calculated cell interfaces does not depend on the values of hydrodynamic quantities in adjacent cells.

### *4.2. Numerical comparisions*

Presence of the term describing the influence of external force at the right-hand side of the equations essentially imposes constrains on to accuracy of computations of fluxes on mesh borders. Furthermore, in the case of fast changing external force, the rate of their formation becomes especially important because it is essential to reflect adequately the influence of external force at the each time step. In order to demonstrate the efficiency of the proposed method we provide the results of numerical experiments that show the rate of formation of adequate fluxes of the mass and the impulse. We compare our results with those obtained by a well-balanced algorithm from [3,9,10,44] for model computations of flow on step..

Initial parameters of the numerical experiments are the following: depth of the fluid is 1 m over the entire spatial area; velocity of the fluid is 5 m/s to the left of the step and 0 m/s to the right of the step (Fig. 3). Step height is 1 m. The evolution of the flux of the mass and the flux of the impulse over the step are displayed on Fig. 4-5.

Black lines indicate the results obtained by the quasi-two-layer method. Grey lines indicate the results given by the well-balanced algorithm from [3,9,10,44]. As one can see, the results obtained by the quasi-two-layer method approaches the appropriate solution in the area of applicability of the method. Howerver the result given by the well-balanced algorithm [3,9,10,44] requires more than 7 time steps before one attaines adequate value of fluxes. For such period of time external force approximated by steps can be changed more the once. Therefore, there may not be enough time to form adequate fluxes during approximated force variations in time.

In the paper [45] it is shown that almost all currently existing methods conserve energy while passing over the step. However, papers [8,45] take into account energy losses. We show below the match between our results with those obtained in [8]. Let us analyze our method in comparison to the well-balanced algorithm in this context [3,9,10,44]. A distinctive feature of the proposed method is the automatic consideration of the properties of the process of the waterfall, i.e. the fluid flow on the step in which the fluid does not wet part of the vertical wall of the step. In this case the dissipation is caused by the lack of interaction with the solid underlying surface for some of the fluid. The work of the gravity force to change the relative depth of the fluid is balanced by the change in the horizontal component of the momentum of the fluid because of this lack. Indicated work is partly used in changing of vertical momentum of the flow that is neglected in the shallow water approximation. Thus the presence of dry zones in the vertical part of the step indicates violation of the conditions of hydrostatic flow. The quasi-two-layer approach determines the size of the dry zone of the vertical component of the step. Consequently it gives an opportunity to figure out the amount of kinetic energy dissipation.

The problem of shock wave propagation over the step has been solved analytically in paper [42]. Initial profile of the shock wave oncoming on the step is showed in Fig. 6. Here $\delta$ is the height of the step, $D_1 = \sqrt{gh_1(h_1+h_2)/2h_2}$ is the velocity of the initial shock wave, $v_1$ is the initial velocity of the fluid to the left of the step and $v_0$ is the initial velocity of the fluid to the right of the step.

The solutions in which the total energy of the flow conserves on the step and the solutions in which the total energy is lost have been considered. The analysis of the problem of the flows which occurs while oncoming of the shock waves on the step has showed that under self-similar solutions of shallow water theory this problem is always solvable. However, the problem does not have a uniquely solution. In paper [42] five qualitatively different types of stable self-similar solutions for this problem have been obtained (Mass fluxes). The total energy of the flow conserves on the step in three of them (solutions types I, III, V). In the other two (solutions types II, IV) the total energy of the flow is lost on the step. The domains of existence of these solutions are plotted on the plane of the dimensionless key parameters (Fig. 7). The subdomains in which both self-similar solutions exist simultaneously have been allocated. Here $h_1^* = h_1/h_0$, $\delta^* = \delta/h_0$, $\alpha \approx 0.66$, $\beta \approx 3.214$.

Unlike most existing methods (see [45]) in which the energy conserves, our approach takes into account the loss of energy while passing over the step in this particular case of shock wave propagation over the step. This is demonstrated in comparison with the results obtained by the well-balancing algorithm [3,9,10,44]. Particularly in Fig. 8 the results of solving the following Riemann problem are shown. The initial flow parameters are the following: to the left from the step the fluid depth is 5.8 m and the velocity is 6 m/s, to the right from the step the fluid depth is 3.3 m and the velocity is 0 m/s and the step height is 1 m. The result obtained by quasi-two-layer method of the second order accuracy is plotted as the black line (solution type II is the rarefaction wave to the left from the step, stationary shock over the step and right shock wave to the right from the step) The result obtained by well-balancing algorithm [3,9,10,44] of the second order accuracy is plotted as grey line (solution type III is the rarefaction wave to the left from the step, stationary shock over the step, left shock wave to the right of the step and right shock wave to the right from the step). In this case there a non-uniqueness of the solution exists for given parameters as shown in paper [42]. Solution obtained by quasi-two-layer method is a solution with energy loss while passing over the step, and solution obtained by well-balancing algorithm [3,9,10,44] is a solution without loss of energy.

Fig. 9 shows the results of solving of the following Riemann problem. The initial flow parameters are as following: to the left from the step the fluid depth is 3 m and the velocity is 8.45 m/s, to the right from the step the fluid depth is 1.5 m and the velocity is 0 m/s and the step height is 1 m. The result obtained by quasi-two-layer method of the second order accuracy is plotted as the black line (solution type IV is stationary shock over the step and right shock wave to the right from the step). The result obtained by well-balancing algorithm [3,9,10,44] of the second order accuracy is plotted as gray line (solution type V is stationary shock over the step, left shock wave to the right from the step and right shock wave to the right from the step). In this case there a non-uniqueness of the solution exists for given parameters as shown in paper [42]. Solution obtained by quasi-two-layer method is a solution with energy loss while passing over the step, and solution obtained by well-balancing algorithm [3,9,10,44] is a solution without loss of energy.

We have also solved fundamental problem of hydrodynamics. Riemann problem on a step has been solved using the quasi-two-layer method. Obtained solutions were compared with the set of exact solutions of the same problem, obtained in paper [8]. In [8] the solutions have been obtained by solving enlarged system that includes an additional equation for the bottom geometry and then the principles of conservation of mass and momentum across the step were used. In order to exclude the multiplicity of solutions, one should impose that the entropy condition is fulfilled; so, total energy dissipates across the stationary shock wave at the step and transition from subcritical to supercritical flow across an upward step is excluded.

The quasi-two-layer method of the second order accuracy is applied for numerical modeling. Fig. 10 shows the solution of the Riemann problem on a step with the following initial conditions: to the left from the step fluid height is 1.462 m, Froud number is 0, to the right from the step fluid height is 0.309 m, Froud number is 0, step height is 0.2 m.

Fig. 11 shows the solution of the Riemann problem on a step with the following initial conditions: to the left from the step fluid height is 0.569 m, Froud number is 0.9, to the right from the step fluid height is 0.569 м, Froud number is 0, step height is 0.2 м.

Good match up between the solutions obtained by the quasi-two-layer method and exact solutions obtained in paper [8] are shown.

## 5. Results of numerical simulations

In order to verify the efficiency of the proposed method a several problems corresponding to different types of nonhomogeneities of the bed and different types of external forces has been solved by using suggested algorithm. In particular, horizontal and sloping planes (section 5.1, 5.4), and a surface of parabolic type were considered as beds (section 5.3). Coriolis force (sections 5.2, 5.3) and a hydraulic friction force (section 5.4) are used for modeling of external forces. The results of numerical computations are compared with available data of laboratory experiments. Interaction of Tsunami waves with the shore line including an obstacle is considered in section 5.5.

## 5.1. Dam-break problem of fluid column of parallelepiped shape on a sloping plane

Let us assume that at the center of an infinite plane, situated at some angle to the horizon, $k$, exist the fluid column of parallelepiped shape. Thus, one considers the effect of the constant external force that is equal to:

$$E = gk \qquad (24)$$

For obtaining the numerical solution, we approximate the inclined plane by the system of ledges with a constant step in space. The selected problem of the fluid column breakdown is most indicative for checking the method efficiency, since it combines in itself all possible types of solutions simultaneously.

The quasi-two-layer method of the first order accuracy is applied for numerical modeling. Comparisons were performed for the cases of resting fluid and fluid flowing upwards along the inclined plane. The spatial step in our computation is taken 0,012 m per cell. The following initial data were taken for comparisons: the size of a discontinuity of the fluid column with height of 1.5 m and square base with side of 2 m over the inclined plane (the inclination value is 1:20), covered with fluid. In the first numerical experiment the fluid was at rest, and in the second one it was flowing onto the plane at velocity of 2 m/s. In both numerical experiments the initial depth was 1 m everywhere outside the region occupied by a breaking column of fluid. The initial flow parameters are shown by a dash-doted line. The dashed line on plots indicates the flow depth and velocity obtained by using the proposed hydrodynamic model, the black line – solutions of combination of two Riemann problems on the slope [26] (Figs. 12, 13).
The presented plots (Figures 12, 13) reveal almost complete analogy between the results obtained by means of the quasi-two-layer method and the solution of the combination of two Riemann problems on the slope [26]. This demonstrates the model efficiency in description of such natural phenomena. Oscillations on the front of hydrodynamic jump occur due to intensity of this hydrodynamic jump. They are within method error and do not result in violating stability with due time.

## 5.2. Rotating fluid flow. The classical Rossby problem

The quasi-two-layer method of the second order accuracy is applied for numerical modeling on space and in time. The rectangular grid of the size 200 X 10 cells is used. The classical Rossby problem is simulated as the test one [46]. The initial disturbance considered was

$$\begin{cases} h(x,0) = h_0 \\ u(x,0) = 0 \\ v(x,0) = Vv_{jet}(x) \end{cases} \qquad (25)$$

where $h_0$ is the initial depth, $V$ is the characteristic scale of velocity, $v_{jet}(x)$ is the normalized profile specified as follows:

$$v_{jet}(x) = \frac{(1+\tanh(4x/L+2))(1-\tanh(4x/L-2))}{(1+\tanh(2))^2}.$$

$L$ is the characteristic scale of disturbance. Characteristic parameters $g, h_0, f$ are fixed. The characteristic scale of velocity $V$ and the characteristic scale of disturbance $L$ are calculated from two dimensionless parameters: the Rossby-Kibel ($Ro$) and Burgers ($Bu$) numbers: $Ro = \frac{V}{fL}$, $Bu = \frac{R_d^2}{L^2}$, where $R_d$ is the deformation radius: $R_d = \frac{\sqrt{gh_0}}{f}$.

The characteristic time scale is specified by the following formula: $T_f = \frac{2\pi}{f}$. The results of evolution of depth $h_0$, in the case of $Ro = 1$, $Bu = 0.25$, are presented below.

Figure 14 shows the evolution obtained by using of the quasi-two-layer model. One can see good coincidence of characteristic peaks of running-away acoustic-gravitational waves and the central balanced part with results presented in paper [10]. This testifies to the efficiency of using the quasi-two-layer model in the description of large-scale geophysical phenomena.

Figure 15 shows the comparison of potential vorticity values at the initial ($t = 0T_f$) and final ($t = 16T_f$) time instants for the classical Rossby problem and $Ro = 1$, $Bu = 0.5$. The potential vorticity is defined the following formula:

$$Q = \frac{\frac{\partial v}{\partial x} + \frac{\partial u}{\partial y} + f}{h} \tag{26}$$

One can see that the invariant $Q$ – the potential vortex is conserved with time. Note that the real time of the process equals to 12 days, approximately. It is seen from the presented plot that the maximum of a function is shifted to the anticyclonic region, and the minimum of the potential vorticity increases with time. The given results are determined by purely nonlinear effects and match well with those obtained in paper [39].

### 5.3. Flow of a rotating fluid over mounted parabolic profile

Problem was simulated, which contained both the nonhomogeneous bed, and the external force – the Coriolis force. The quasi-two-layer method of the second order accuracy is applied for numerical modeling on space and time. The shallow water equations in this case are as follows:

$$\begin{cases} \dfrac{\partial h}{\partial t} + \dfrac{\partial (hu)}{\partial x} + \dfrac{\partial (hv)}{\partial y} = 0 \\ \dfrac{\partial (hu)}{\partial t} + \dfrac{\partial \left(hu^2 + gh^2/2\right)}{\partial x} + \dfrac{\partial huv}{\partial y} = -gh\dfrac{db}{dx} + fvh \\ \dfrac{\partial (hv)}{\partial t} + \dfrac{\partial \left(hv^2 + gh^2/2\right)}{\partial y} + \dfrac{\partial hvu}{\partial x} = -gh\dfrac{db}{dy} - fuh \end{cases} \qquad (27)$$

where $f$ is the Coriolis parameter. According to (27), we get the external forces: $E_x = fv$ and $E_y = -fu$. The large-scale motion of fluid in the gravitational field in the presence of Coriolis force over the mountain-like bed is considered. Typical parameters of the problem are: the linear dimensions $10^6$ by $10^6$ m, the mountain height is $1,2 \cdot 10^3$ m, the fluid (depth is $2 \cdot 10^3$ m, and the Coriolis parameter is 0.00001452 s$^{-1}$. The initial wind parameters are $u = 0$ m / s, $v = 20$ m / s. The boundary free-slip conditions are satisfied. The computational domain contains 3600 cells (60×60 cells). According to (5) the external forces $E_x = fv$ and $E_y = -fu$, the corresponding differences of average heights of a fictitious surface on the boundary of the two adjacing cells are: $\dfrac{\left|E_{x_r}X/g + E_{x_l}X/g\right|}{2} = \dfrac{\left|fv_r X/g + fv_l X/g\right|}{2}$, $\dfrac{\left|E_{y_r}X/g + E_{y_l}X/g\right|}{2} = \dfrac{\left|-fu_r X/g - fu_l X/g\right|}{2}$.

As a result, we found that the characteristic time of one revolution of a system as a whole is 24 hours, which corresponds to the natural phenomenon (the characteristic time of one revolution of a system as a whole for the geophysical dynamics problems equals to one day [16]). Figure 16 shows the picture of flow at the initial time instant; Figure 17 shows the flow evolution during 24 hours.

### 5.4. Two-dimensional dam-break problem taking into account a hydraulic friction

The computational domain represents a rectangle of size 3 X 2 m, restricted by non-leaking walls on three sides. There is a wall apart 1 m from the left non-leaking boundary with a hole of width 0.40 m which is symmetrical with respect to the abscissa axis. The thickness of the wall is a negligible quantity and it is not used in calculations. A hole in a wall is initially closed and all the fluid rests on the left of it. The fluid can freely leak through the right boundary of the computational domain. Define the Cartesian coordinate system in the following manner: the wall belongs to the ordinate axis, and the abscissa axis divides the computational domain in two symmetrical parts. The origin of coordinates coincides with the centre of the hole in a wall. All requirements of the performed numerical modeling are taken according to requirements of laboratory and numerical experiments from [17]. The slope angle of a bed in the performed numerical and laboratory experiments varied from 0 % to 10 %. Coordinates of control points and the plan of their location are shown on Fig. 18. Laboratory installation is made from plexiglass and supplied by the mechanism for opening a shutter of the hole in the wall at a velocity sufficient that dynamics of uprise of the shutter was not reflected in a pattern of formed flow. Detailed description of laboratory installation and the numerical experiments can be fined in the paper of Fraccarollo and Toro [17].

The shallow water equations in this case are follows:

$$\begin{cases} \dfrac{\partial h}{\partial t} + \dfrac{\partial (hu)}{\partial x} + \dfrac{\partial (hv)}{\partial y} = 0 \\ \dfrac{\partial (hu)}{\partial t} + \dfrac{\partial \left(hu^2 + gh^2/2\right)}{\partial x} + \dfrac{\partial huv}{\partial y} = -gh\dfrac{db}{dx} + \dfrac{1}{2}\lambda u|u|h \\ \dfrac{\partial (hv)}{\partial t} + \dfrac{\partial \left(hv^2 + gh^2/2\right)}{\partial y} + \dfrac{\partial hvu}{\partial x} = -gh\dfrac{db}{dy} + \dfrac{1}{2}\lambda v|v|h \end{cases} \qquad (28)$$

where $\lambda = 2gn^2 h^{-\frac{4}{3}}$, $n = 0.007$. External force for system (28) have the form of hydraulic friction: $E_x = \dfrac{1}{2}\lambda u|u|$, $E_y = \dfrac{1}{2}\lambda v|v|$. According (5) the corresponding differences of average heights of a fictitious surface on the boundary of the two neighboring cells are:

$$\frac{\left|E_{x_r} X/g + E_{x_l} X/g\right|}{2} = \frac{\left|\frac{1}{2}\lambda u_r |u_r| X/g + \frac{1}{2}\lambda u_l |u_l| X/g\right|}{2},$$

$$\frac{\left|E_{y_r} X/g + E_{y_l} X/g\right|}{2} = \frac{\left|\frac{1}{2}\lambda v_r |v_r| X/g + \frac{1}{2}\lambda v_l |v_l| X/g\right|}{2}.$$

The quasi-two-layer method of the second order accuracy is applied for numerical modelling on space and time. Results for the case of horizontal and sloping beds are shown below. The slope angle is 6.3 degrees. Depth of a fluid to the left of a wall is 0.6 м at initial time moment, the fluid was at rest for all the results given below. The computational domain is divided by regular grid of size 150 × 50.

Plots of fluid depths and velocities in a point 0 corresponding to the centre of dam-break are shown on Fig 19. The left plot corresponds to fluid depths, right – fluid velocities. Thin grey line corresponds to the data obtained in laboratory experiment [17], dashed grey line - to numerical results of WAF method [17], heavy black line – to numerical results of proposed quasi-two-layer method.

Dynamics of fluid depth for a horizontal bed in control point P1 is shown on Fig. 20: obtained experimentally – thin grey line, and numerically on the basis of WAF method [17] – dashed grey line, heavy black line – on the basis of proposed quasi-two-layer method, thin black line – on the basis of proposed quasi-two-layer method without bed friction. One can see that numerical results obtained without bed friction are slightly different from those obtained with bed friction in control point P1. Control point P1 is located very closely to localization domain of the whole initial mass of the fluid. In this particular case bed friction is important. Indeed an integral bed friction effect of the whole domain where the fluid spread affect on flow parameters in control point P1 in spite of low bed friction effect in local control point P1. Note that this integral effect of bed friction on flow parameters is the maximal distinctive in domain where the whole initial mass of the fluid is. This total effect decreases as the control point moves away from this domain and becomes negligible.

Dynamics of fluid depth for a horizontal bed in two control points are shown on Figs. 21-22: obtained experimentally – thin grey line, and numerically on the basis of WAF method [17] – dashed grey line, heavy black line – on the basis of proposed quasi-two-layer method.

Dynamics of fluid depth for a sloping bed in two control points are shown on Fig. 23. Thin lines correspond to the data obtained in laboratory experiment: dashed – point P2,

solid – point 0. Numerical results of proposed quasi-two-layer method shown by heavy lines: dashed – point P2, solid – point 0. On Fig. 24 numerical results of WAF method for the same points are shown. One can see that proposed quasi-two-layer method is in good agreement with experimental data and improves accuracy of results in comparison with the numerical results of WAF method obtained in article [17].

*5.5. Tsunami waves simulation*

Interaction of the Tsunami wave with the shore line with and without an obstacle has been simulated to demonstrate the effectiveness of the developed algorithm in domains including partly flooded and dry regions. The computational domain is divided into regular grid of size 1200 × 1. Slope angle of the bed is 0.125. Initial height of the Tsunami wave is 12 m, velocity is 5 m/s. Interaction of the Tsunami wave with the shore line is shown on Fig. 25. Dynamics of fluid depth in various time steps (indicated on the wave front in seconds) is shown on the upper figure. Dynamics of fluid velocity in various time steps (indicated on the wave front in seconds) is shown on the lower figure. The significant increase of fluid velocity and Zunami wave propagation for the distance about 900 m is observed during interaction of the Tsunami wave with the shore line.

Interaction of the Tsunami wave with the shore including an obstacle of size 12 m located 100 m away from the coast has been simulated. Dynamics of fluid depth in the various time steps (indicated on the wave front in seconds) is shown on Fig. 26. Dynamics of fluid velocity in various time steps (indicated on the wave front in seconds) is shown on Fig. 27. The considerable decrease of fluid velocity after passing an obstacle and Zunami wave propagation for the distance about 750 m is observed. The decrease of destroying effect of the Tsunami wave on the shore line by an obstacle is shown. Scaled-up area with an obstacle on the shore is shown on Fig. 28. On the left is the dynamics of the fluid depth in various time steps (indicated on the obstacle in seconds). On the right is the Dynamics of the fluid velocity in the various time steps. The reflected shock wave propagating away from the obstacle and some of the fluid flowing around the obstacle are observed. Thus effectiveness of the developed algorithm in domains including partly flooded areas and dry regions is demonstrated.

**6. Conclusion.**

New numerical method for modeling of shallow water flows over an arbitrary bed profile in the presence of external force is proposed. The presentation of arbitrary external force by a fictitious bed along with the bed relief features is approximated by a time-dependent stepwise bed. This allows to apply the finite volume methods based on the Riemann problem solution for determination the flow quantities. To the Riemann problem we use an advanced algorithm which is based on the two-layer representation of fluid flow near the step. Such representation allows to obtain in suggested new finite difference scheme the additional terms that provide mechanical work made by nonhomogeneous bed interacting with flow. This also modifies properly values of hydrodynamic fluxes in steps vicinity. The calculation of flows in the proposed algorithm is performed on the basis of the quasi-two-layer shallow water model, which represents the generalization of the classical single-layer model in relation to the initial system of Euler equations. It is shown that the developed finite-difference scheme belongs to the well-balance class. Unlike the known schemes of well-balanced type, the proposed scheme is adapted to flow parameters. The developed approach allows one to restore the flow structure throughout the space-time domain with consideration for its vertical peculiarities near the step. This makes it possible to determine the transversal component of velocity vector more correctly for the class of external forces, which determine essentially two-dimensional problems, such as the problems with the Coriolis force. We will benefit this quality in future developments. The calculation of essentially two-dimensional problems with a corrected transversal velocity component will be accomplished in a separate work. A distinctive feature of the proposed method is the consideration of the properties of the process of the waterfall, i.e. the fluid flow on the step in which the fluid does not wet part of the vertical wall of the step. The presence of dry zones in the vertical part of the step indicates violation of the conditions of hydrostatic flow. The quasi-two-layer approach allows to determine the size of the dry zone of the vertical component of the step. Consequently it gives an opportunity to figure out the amount of kinetic energy dissipation. In order to demonstrate the efficiency of the proposed method we have demonstrated results of the computations that show the rate of formation of an adequate flux of the mass and the impulse. We compare our results with those obtained by the well-balancing method. The Riemann problem on a step has been solved and good agreement of the obtained solutions and exact solutions is shown for non-waterfall case. The applicability of the method is demonstrated by the example of solution of the problem of breakdown of a moving fluid column on an inclined plane. The motion of fluid in the presence of Coriolis force over a bed of the given profile is simulated. Reliability and validity of our results is shown on the basis of the solution of two-dimensional dam-break problem taking into account a hydraulic friction and comparison of the obtained solutions with available laboratory experiments results. Interaction of the Tsunami wave with the shore line including an obstacle has been simulated to demonstrate the effectiveness of the developed algorithm in domains with partly flooded and dry regions.

**Acknowledgements**

The authors are grateful to Rupert Klein for reading manuscript and useful comments.

**Figure captions**

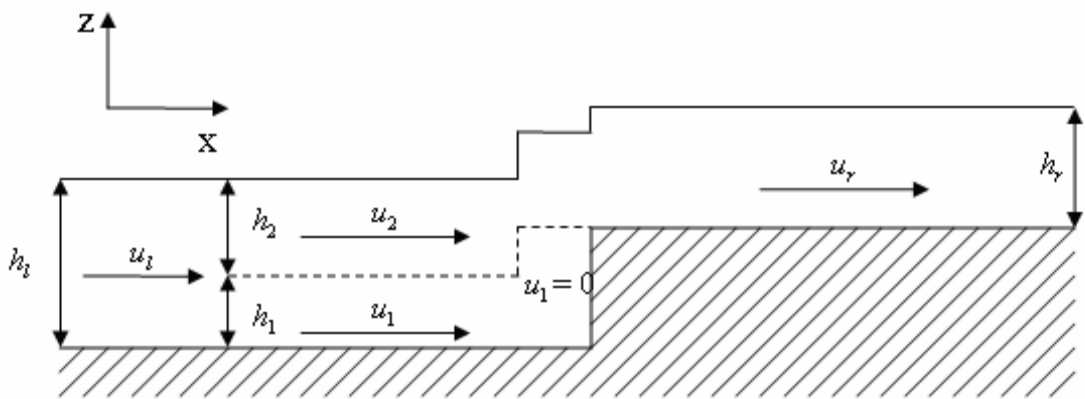

Fig. 1. Quasi-two layer representation of shallow flow on step

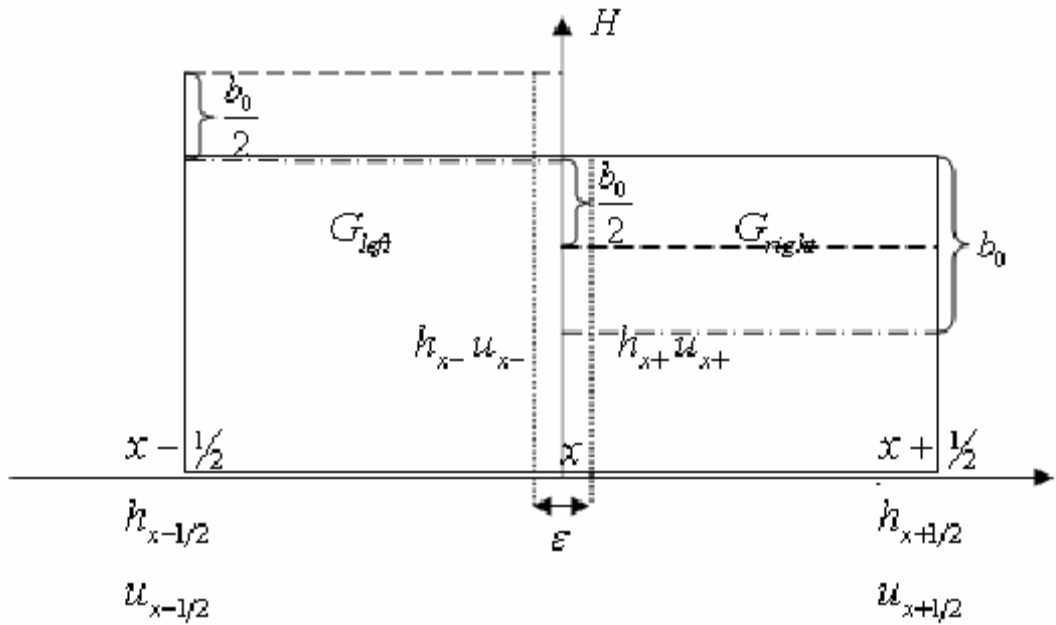

Fig. 2. Scheme of updating flow parameters in the Wave propagation [41,42] method (dashed line) and in the well-balancing [3,9,10,44] method (dash-dotted line)

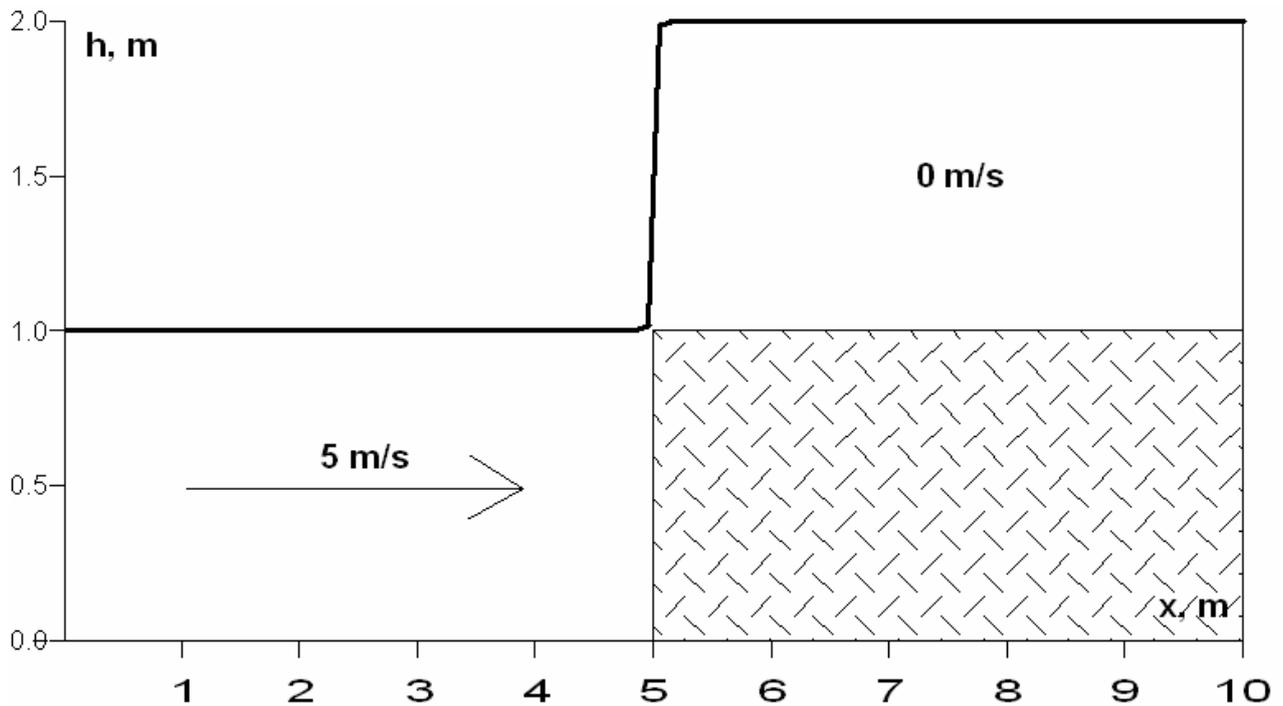

Fig. 3. Initial parameters of the fluid flow for different finite difference schemes comparisions

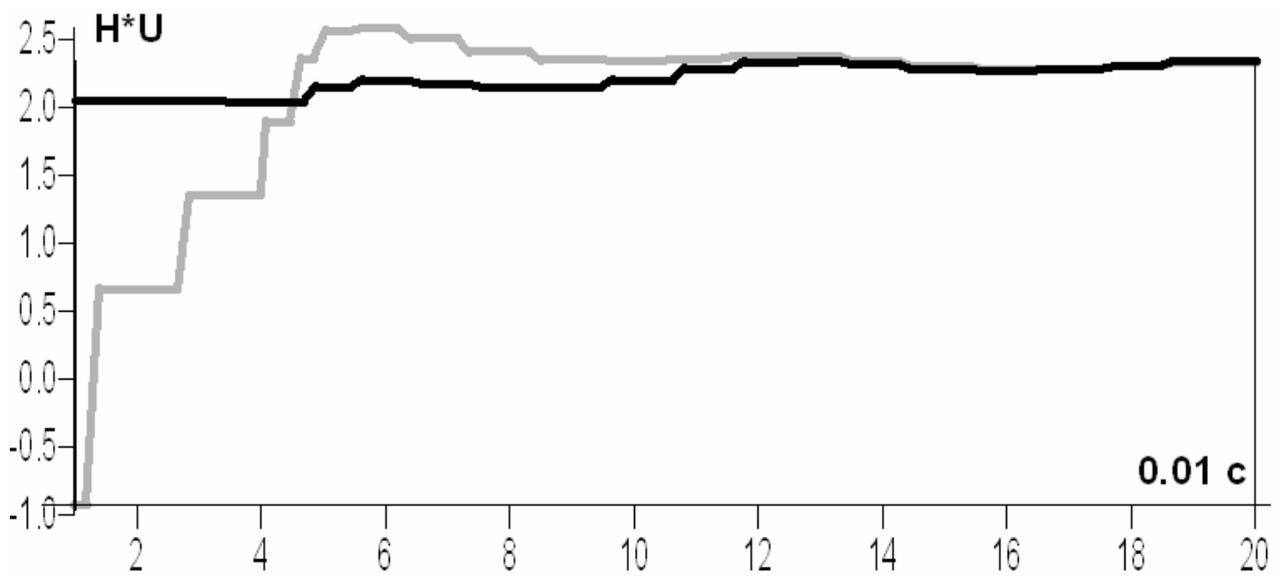

Fig. 4. Comparisions for the mass fluxes in time over the step. Black lines indicate the results obtained by the quasi-two-layer method. Grey lines indicate the results given by the well-balancing algorithm.

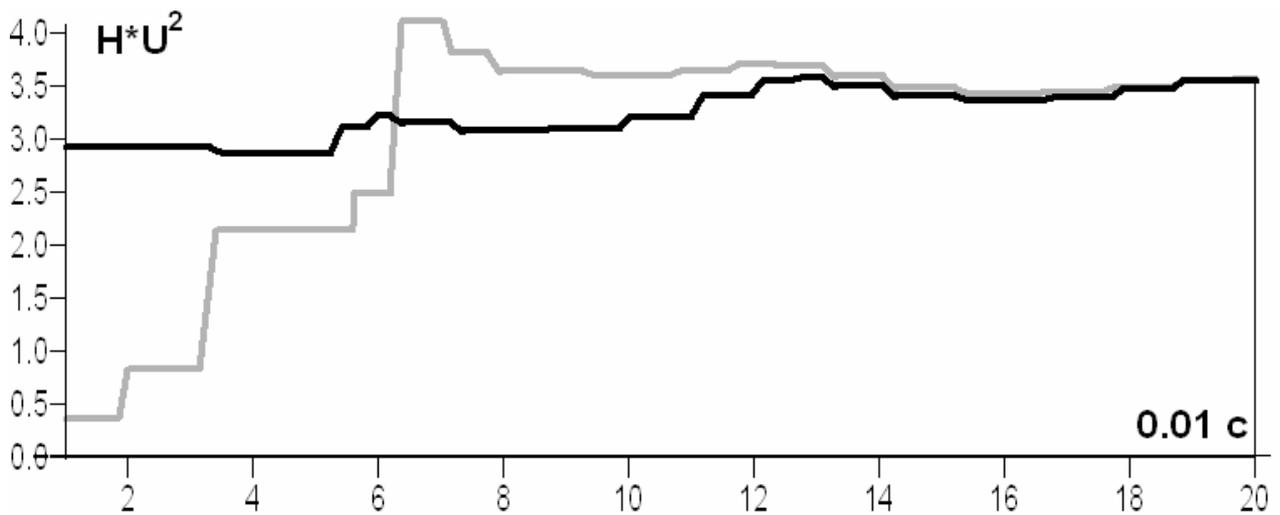

Fig. 5. Time evolution of the flux of the impulse over the step. Black lines indicate the results obtained by the quasi-two-layer method. Grey lines indicate the results given by the well-balancing algorithm.

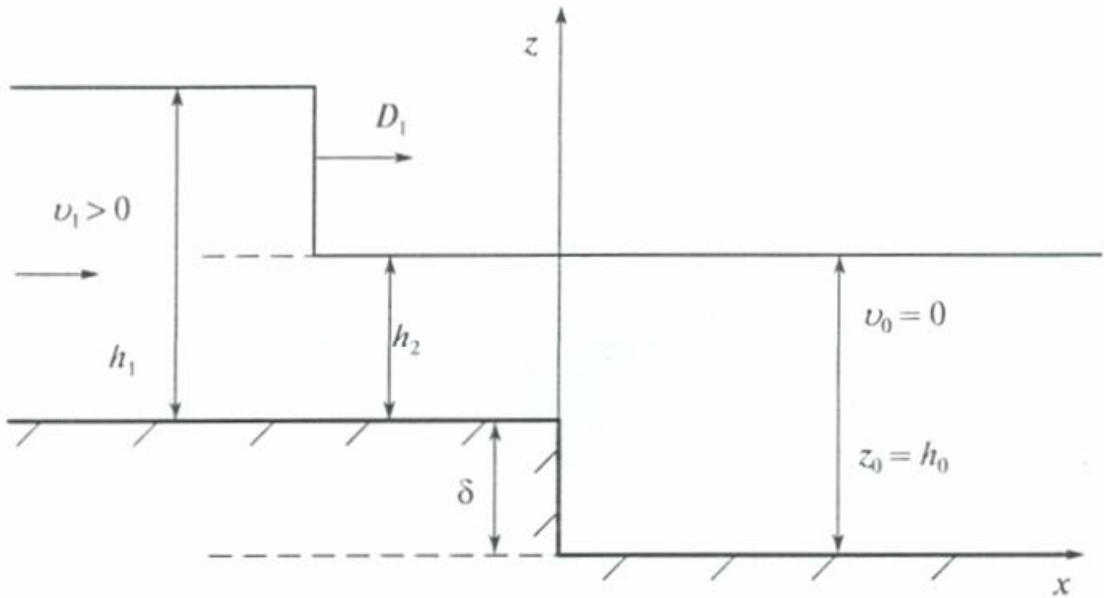

Fig. 6. Initial condition of the shock wave oncoming on the step (from paper [42])

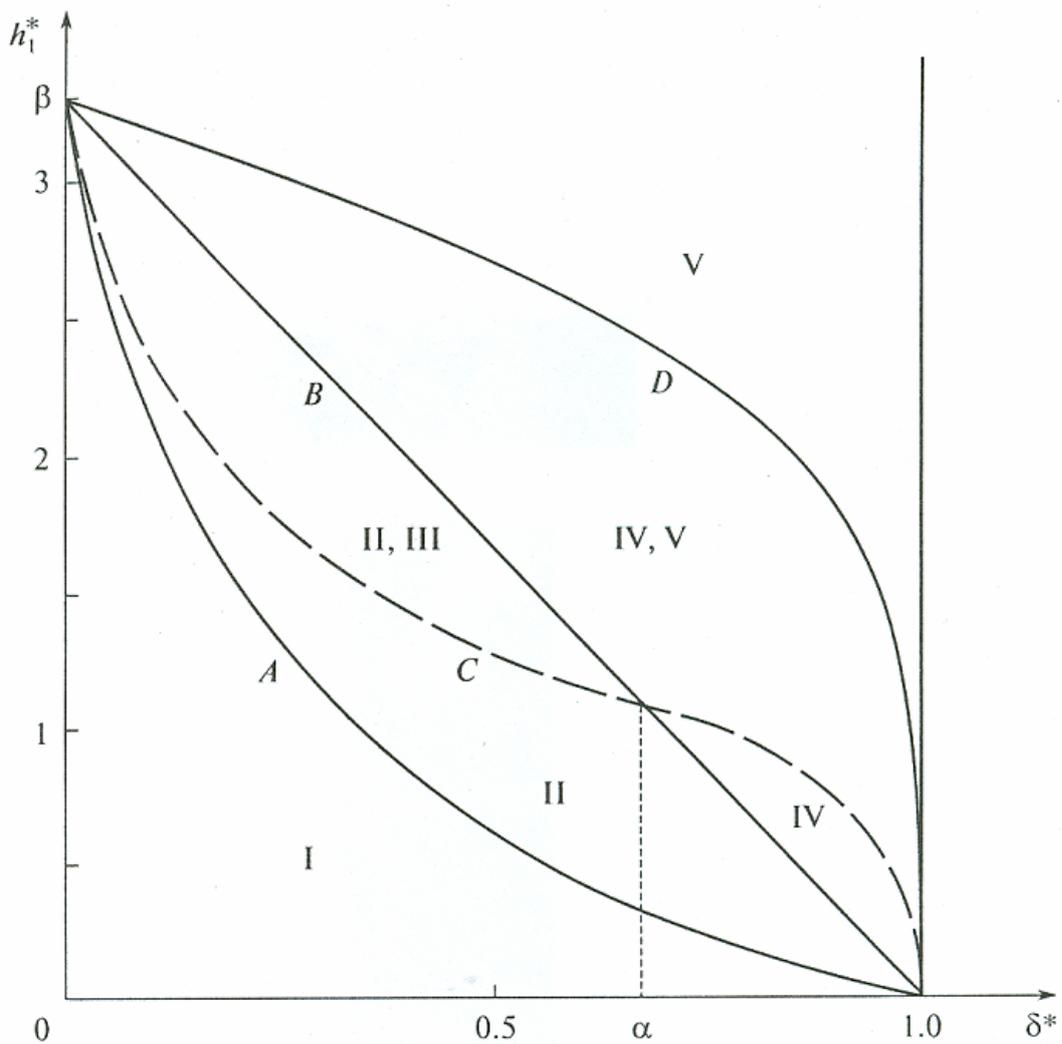

Fig. 7. Diagram of the dimensionless key parameters characterizing non-uniqueness of flow on step (from paper [42])

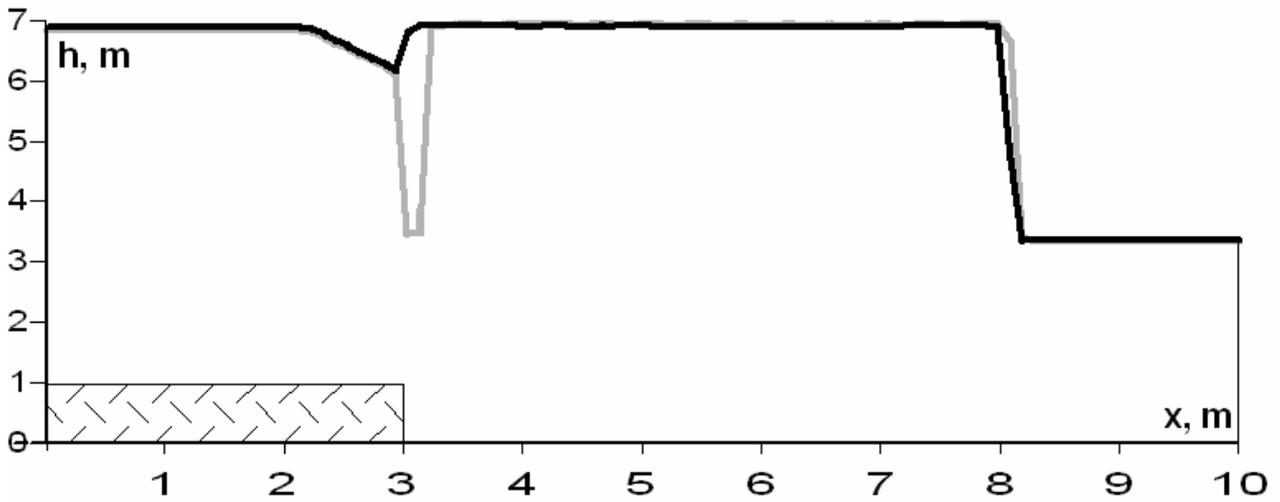

Fig. 8. Fluid depths during propagation of the flow through the step at time 0.4 s. Black lines indicate the results obtained by the quasi-two-layer method. Grey lines indicate the results given by the well-balancing algorithm.

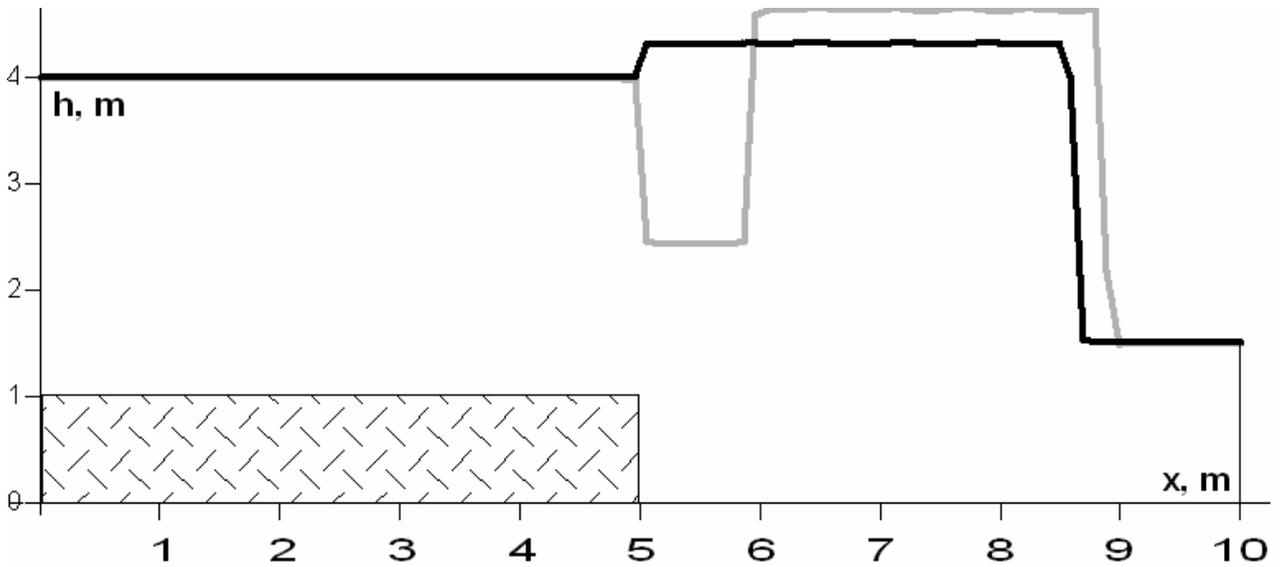

Fig. 9. Fluid depths during passage of the flow through the step at time 0.4 s. Black lines indicate the results obtained by the quasi-two-layer method. Grey lines indicate the results given by the well-balancing algorithm.

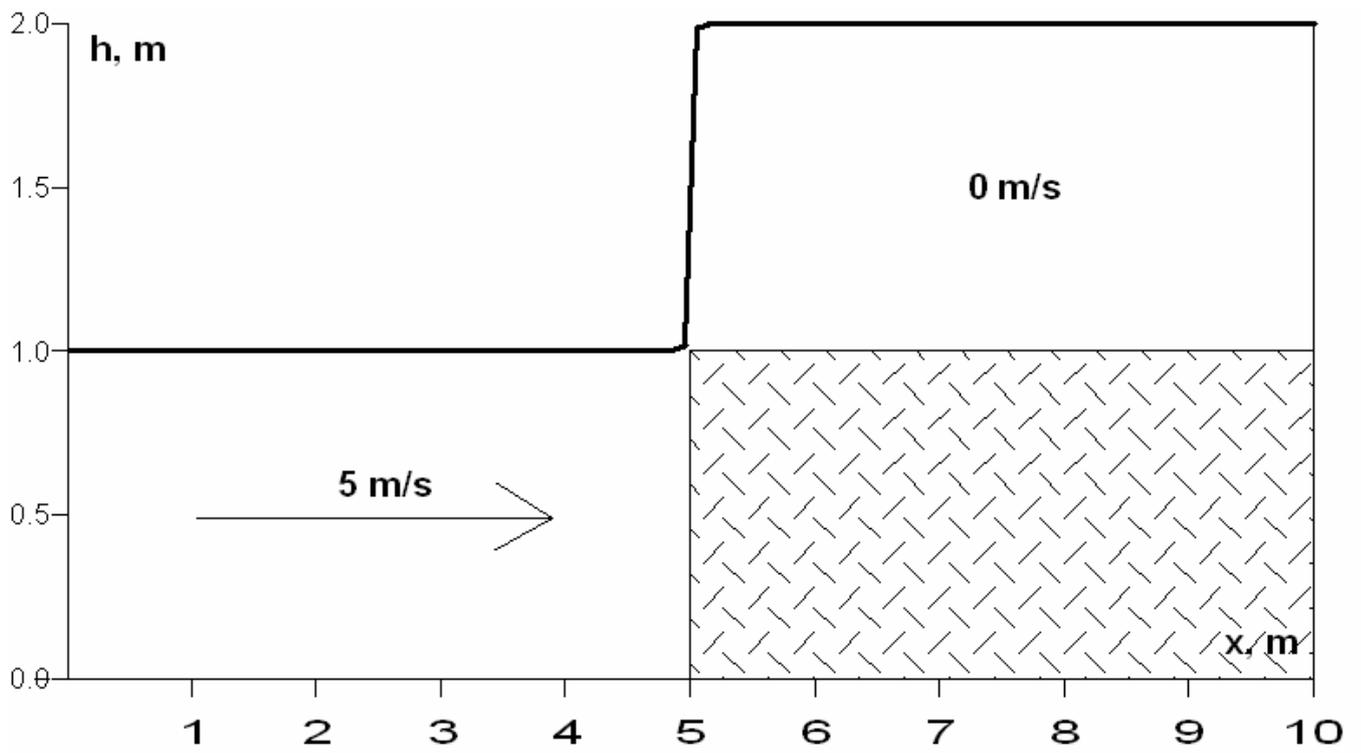

Fig. 10. Comparison of quasi-two-layer solution (left) with the exact solution (right). Configuration: the left rarefaction wave to the left from the step, the right shock to the right from the step.

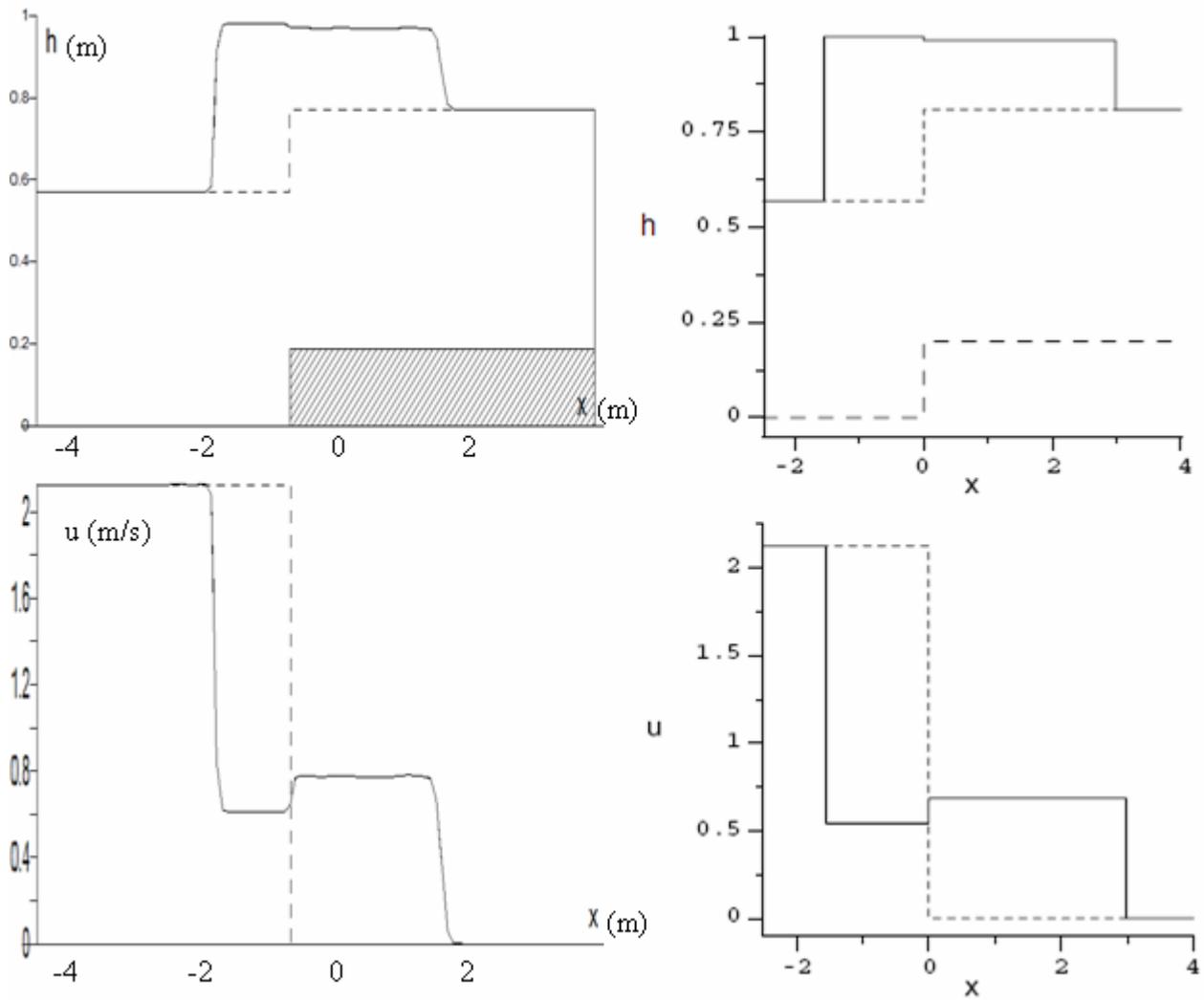

Fig. 11. Comparison of the quasi-two-layer solution (left) with the exact solution (right). Configuration: the left shock to the left from the step, the right shock to the right from the step.

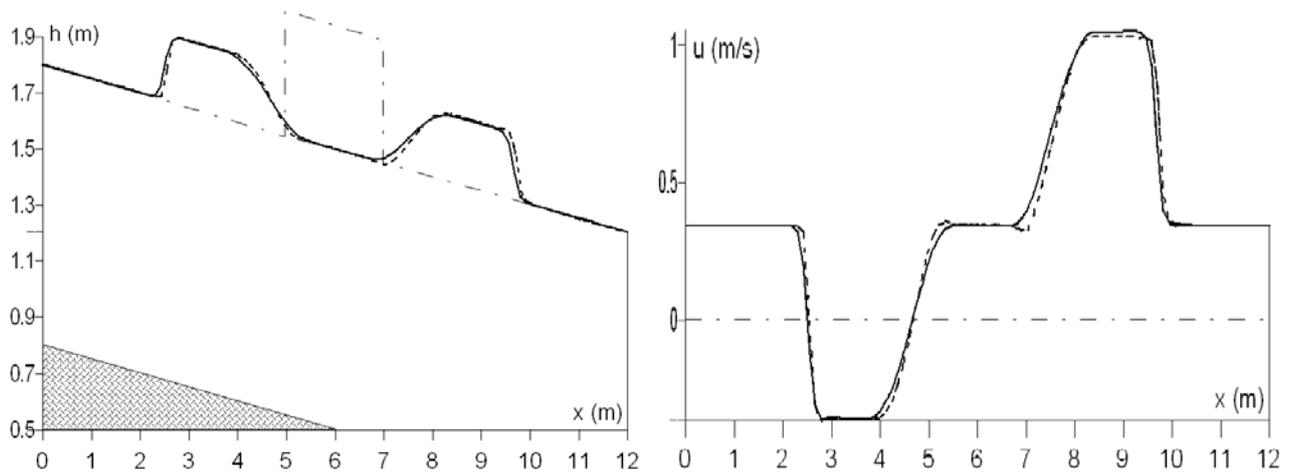

Fig. 12. Depth and velocity of fluid flow over an inclined bed, $t$ = 0.7 s in the cross section by the plane of symmetry, $y$ = 0. Dash-doted line - initial flow parameters; dashed line -

flow depth and velocity obtained by using the proposed hydrodynamic model; black line – flow depth and velocity obtained by decision of the Riemann problems on the slope

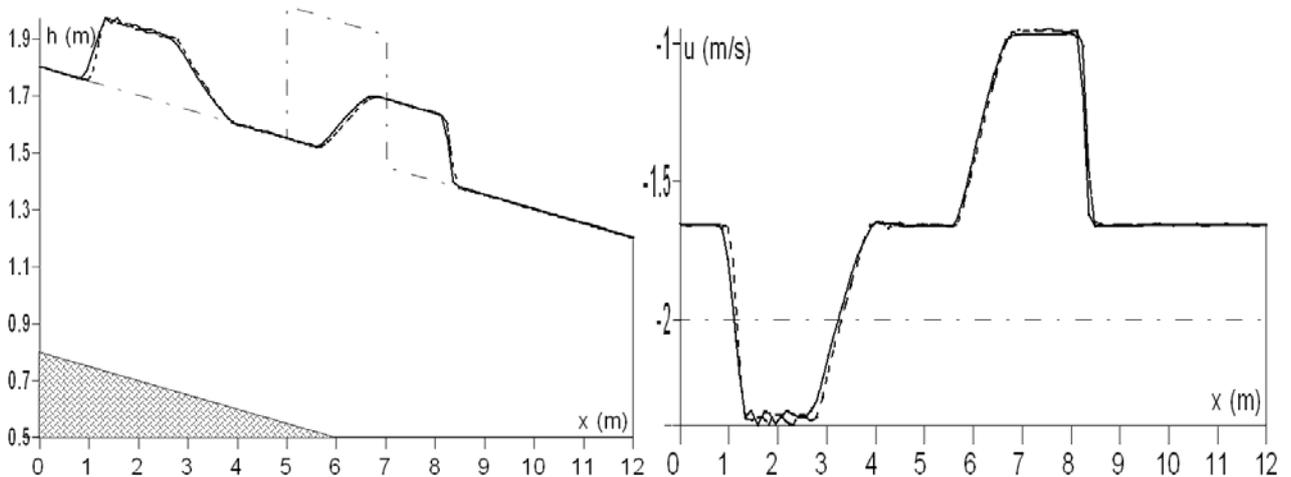

Fig. 13. Depth and velocity of fluid flow over an inclined bed with fluid leakage, $t$ = 0.7 s in the cross section by the plane of symmetry, $y$ = 0. Dash-doted line - initial flow parameters; dashed line - flow depth and velocity obtained by using the proposed hydrodynamic model; black line – flow depth and velocity obtained by decision of the Riemann problems on the slope

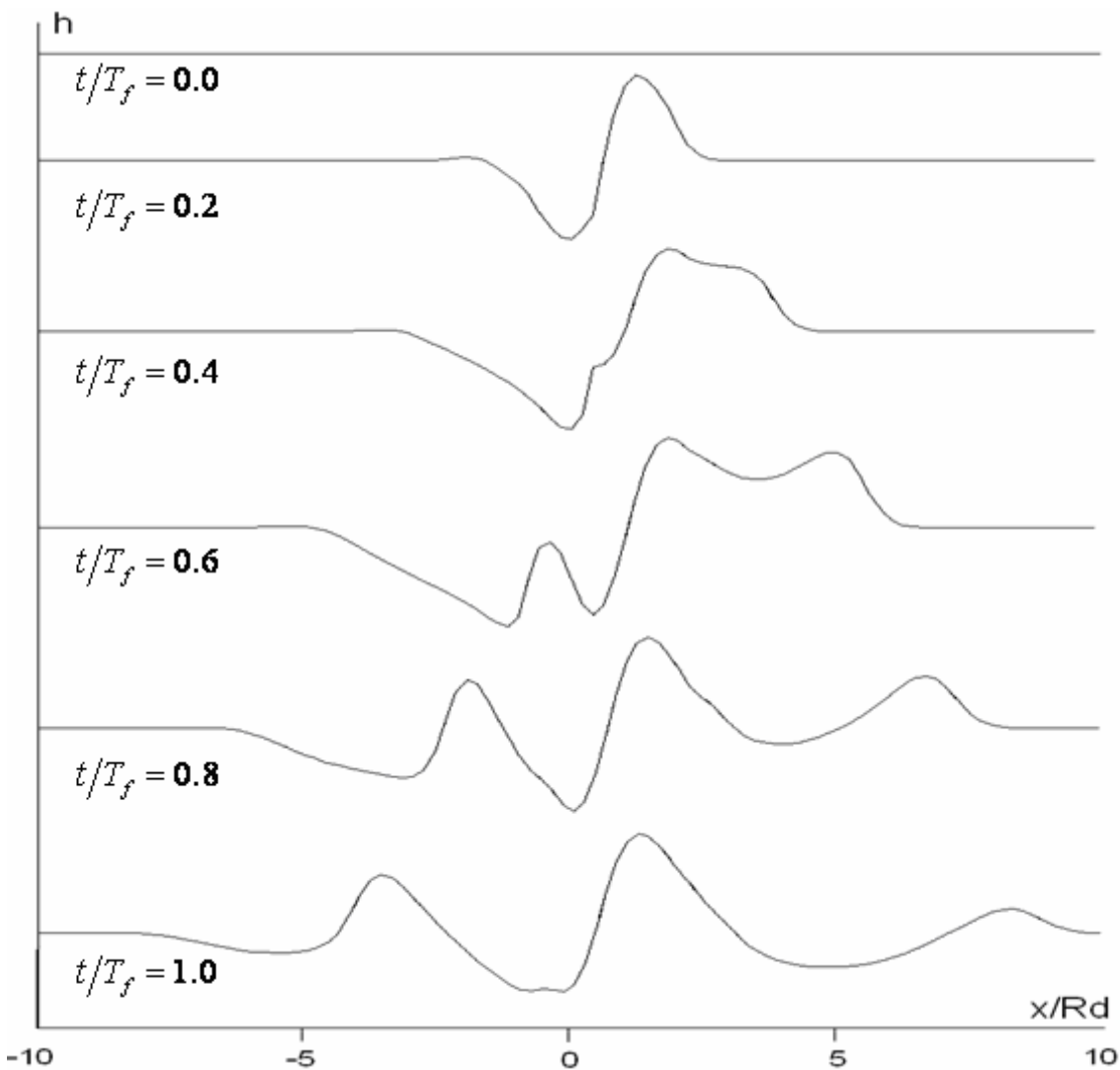

Fig. 14. Results of computations of propagation of acoustic-gravitational waves using the quasi-two-layer method

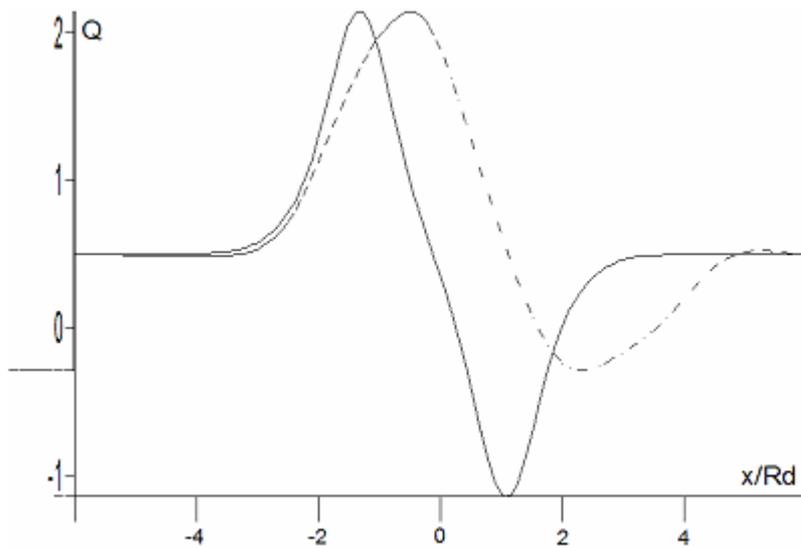

Fig. 15. Potential vorticity computations using quasi-two-layer method at the initial $t = 0,2T_f$ (black line) and at the final $t = 16T_f$ (dashed line).

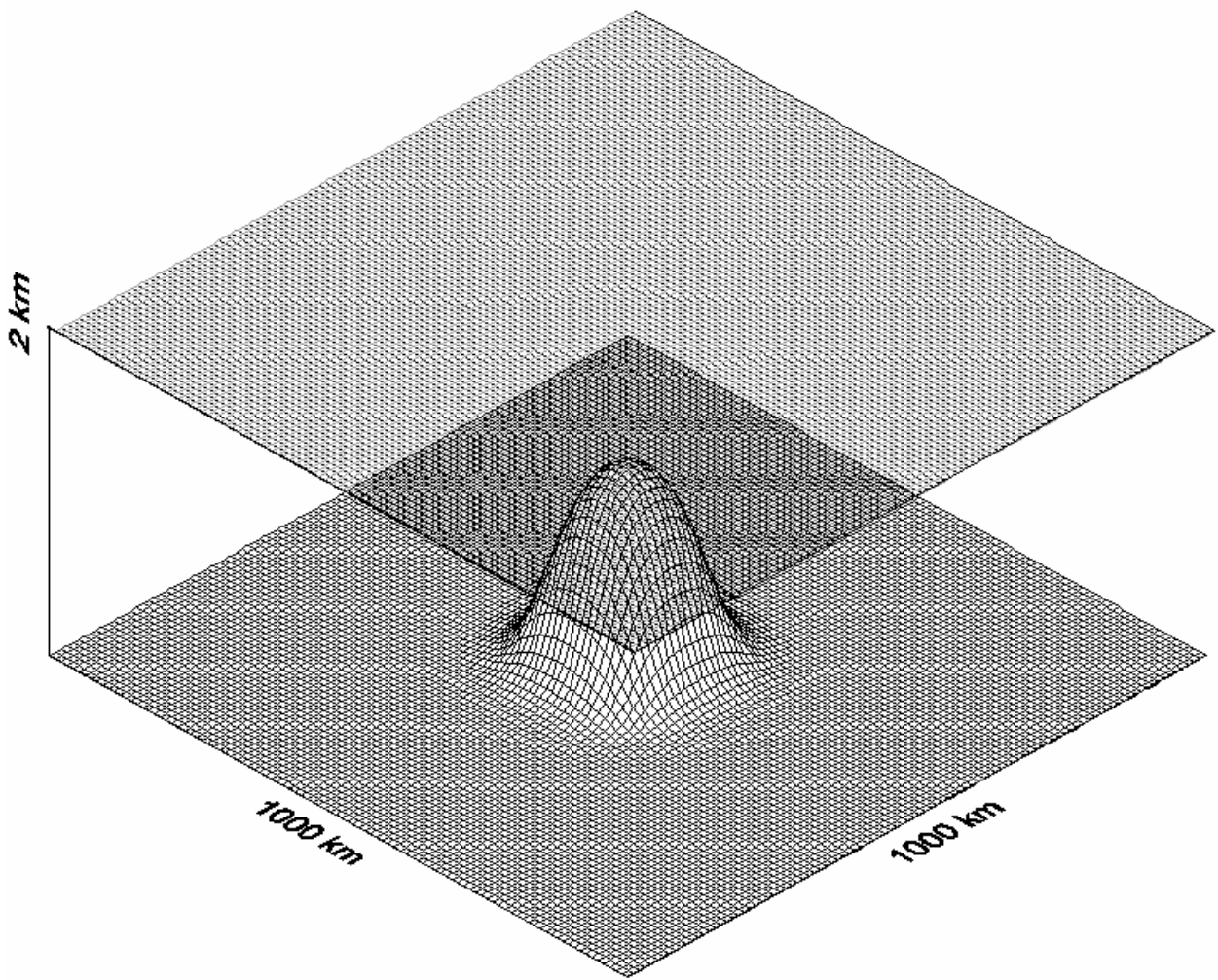

Fig. 16. Initial conditions of the test problem of the flow of rotating fluid over mounted parabolic profile.

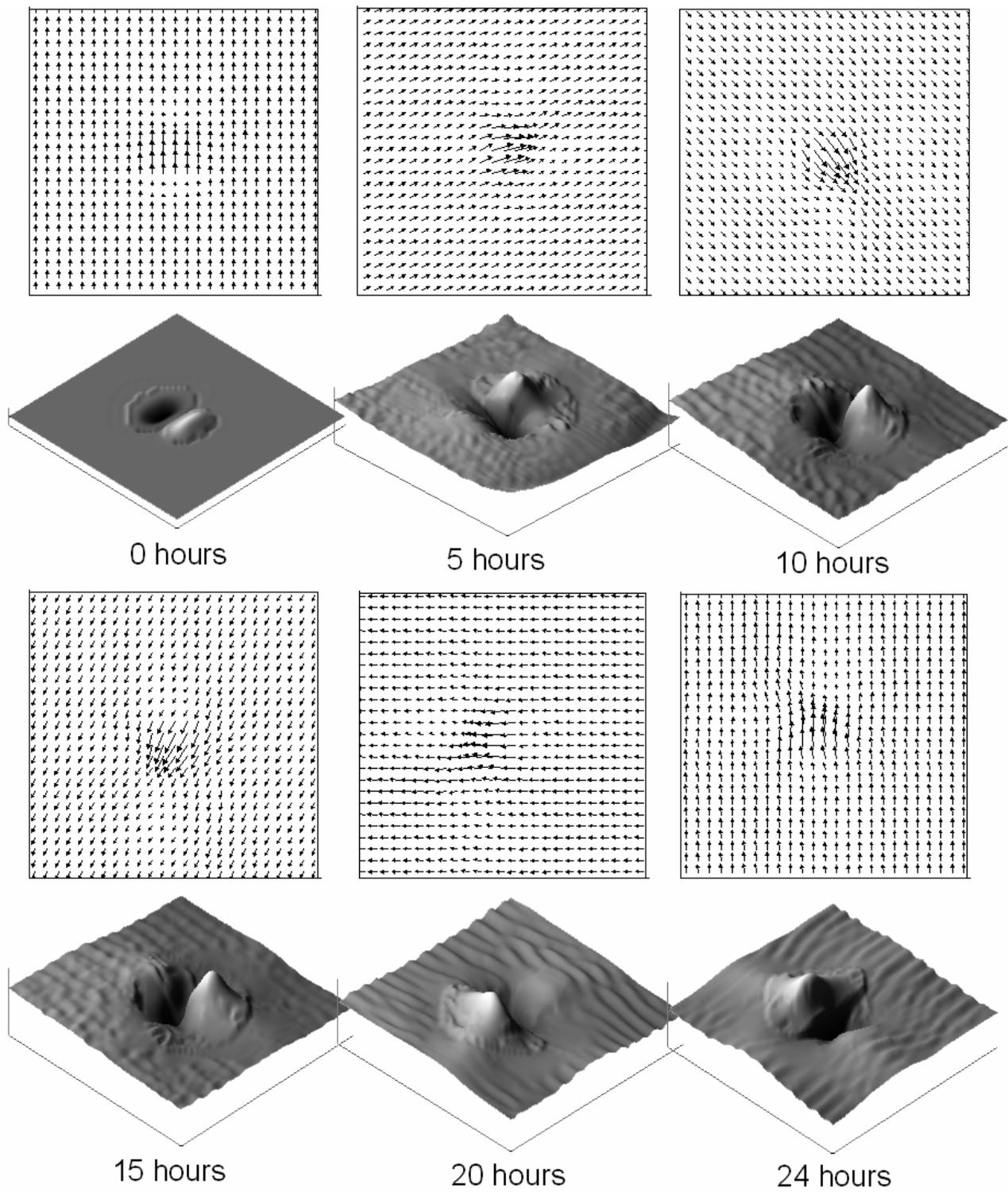

Fig. 17. Evolution of fluid (gas) flow under the Coriolis force effect over a mountain. Upper plots are velocity fields, lower plots are the free surface depth

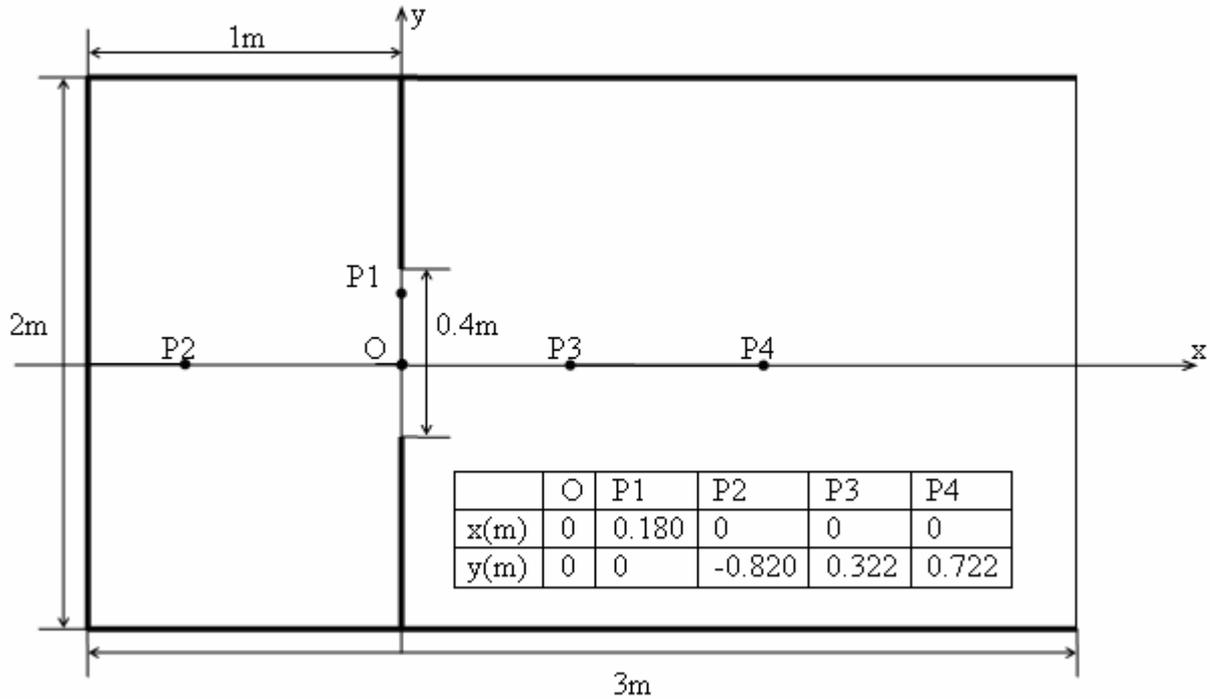

Fig. 18. Computational domain and coordinates of control points of two-dimensional dam-break problem calculations

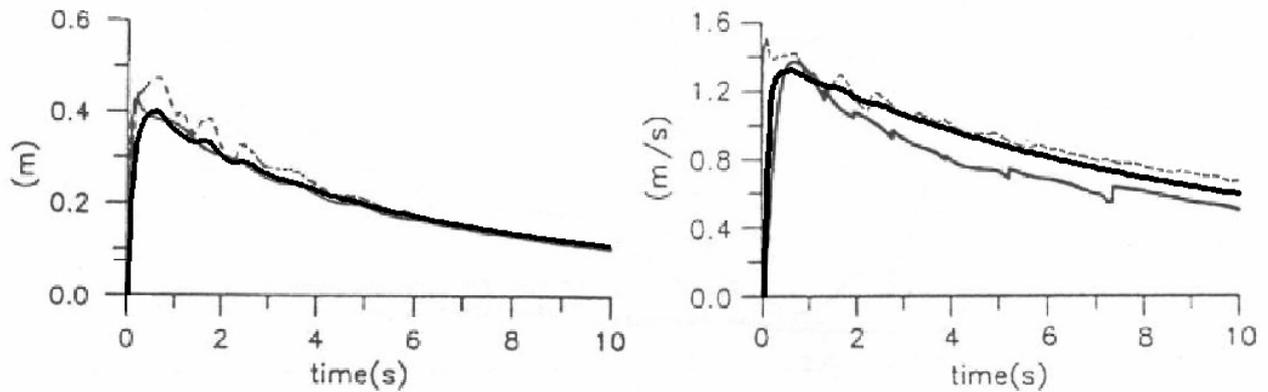

Fig. 19. Left: fluid depths on time at point O. Right: fluid velocities on time at point O. Thin grey line - data obtained in laboratory experiment, dashed grey line - numerical results of WAF method, heavy black line – numerical results of proposed quasi-two-layer method

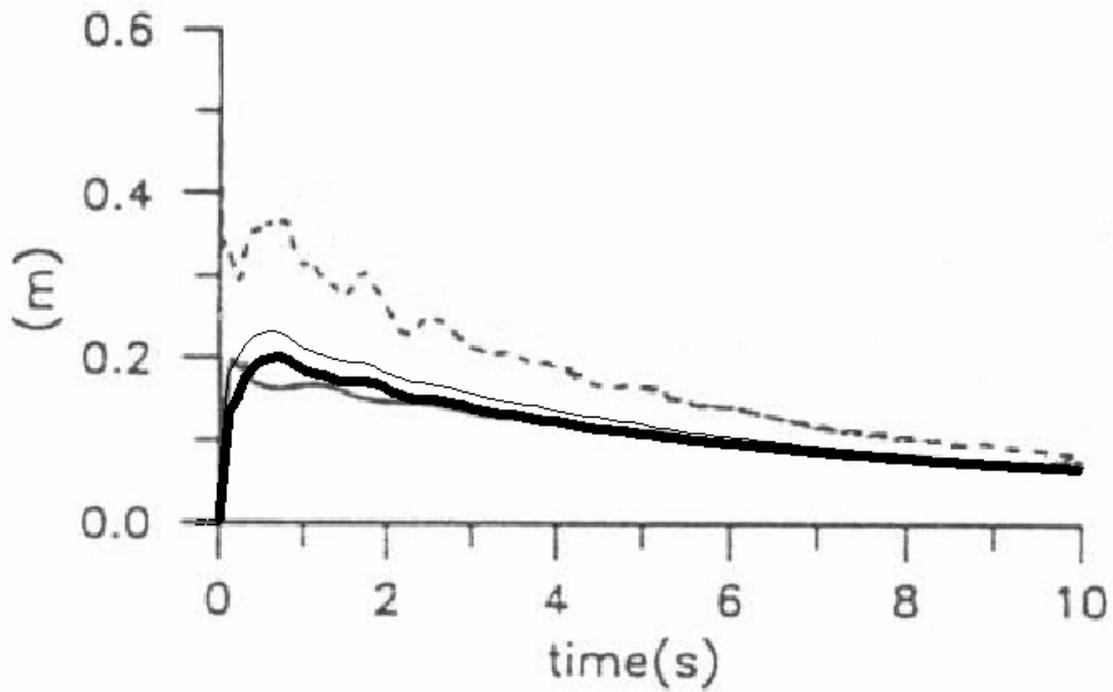

Fig. 20. Fluid depths on time at point P1. Thin grey line - data obtained in laboratory experiment, dashed grey line - numerical results of WAF method, heavy black line – numerical results of proposed quasi-two-layer method

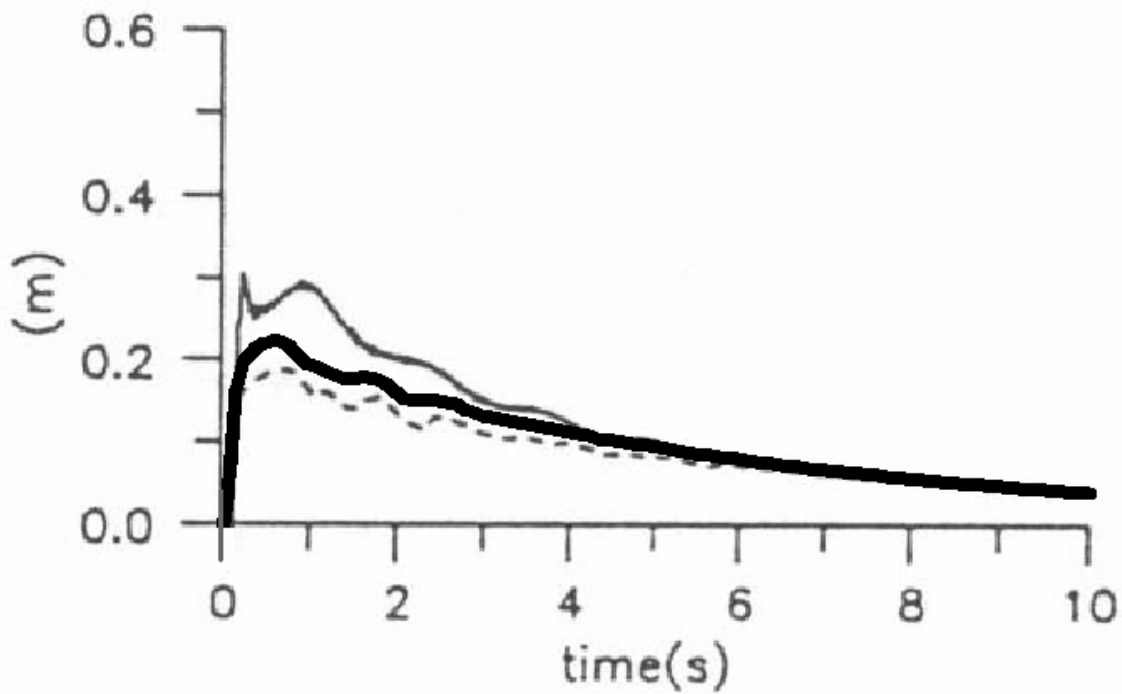

Fig. 21. Fluid depths on time at point P3. Thin grey line - data obtained in laboratory experiment, dashed grey line - numerical results of WAF method, heavy black line – numerical results of proposed quasi-two-layer method

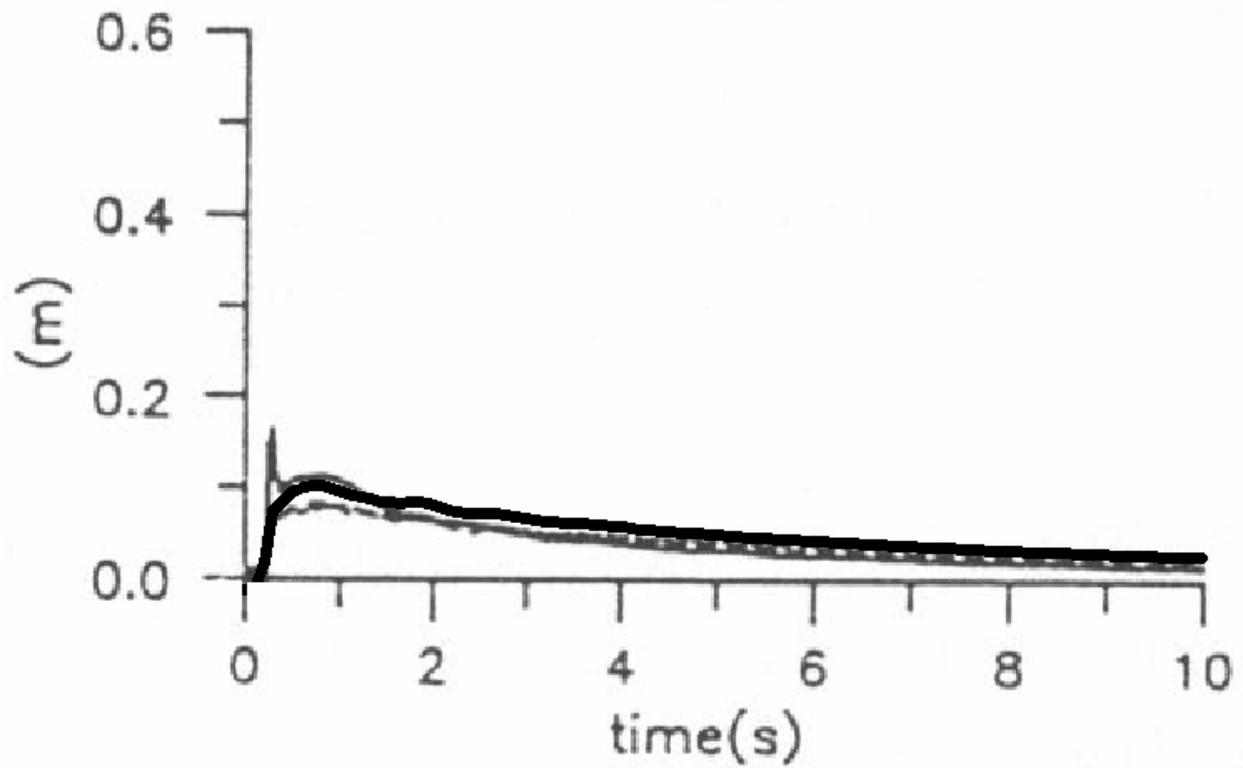

Fig. 22. Fluid depths on time at point P4. Thin grey line - data obtained in laboratory experiment, dashed grey line - numerical results of WAF method, heavy black line – numerical results of proposed quasi-two-layer method

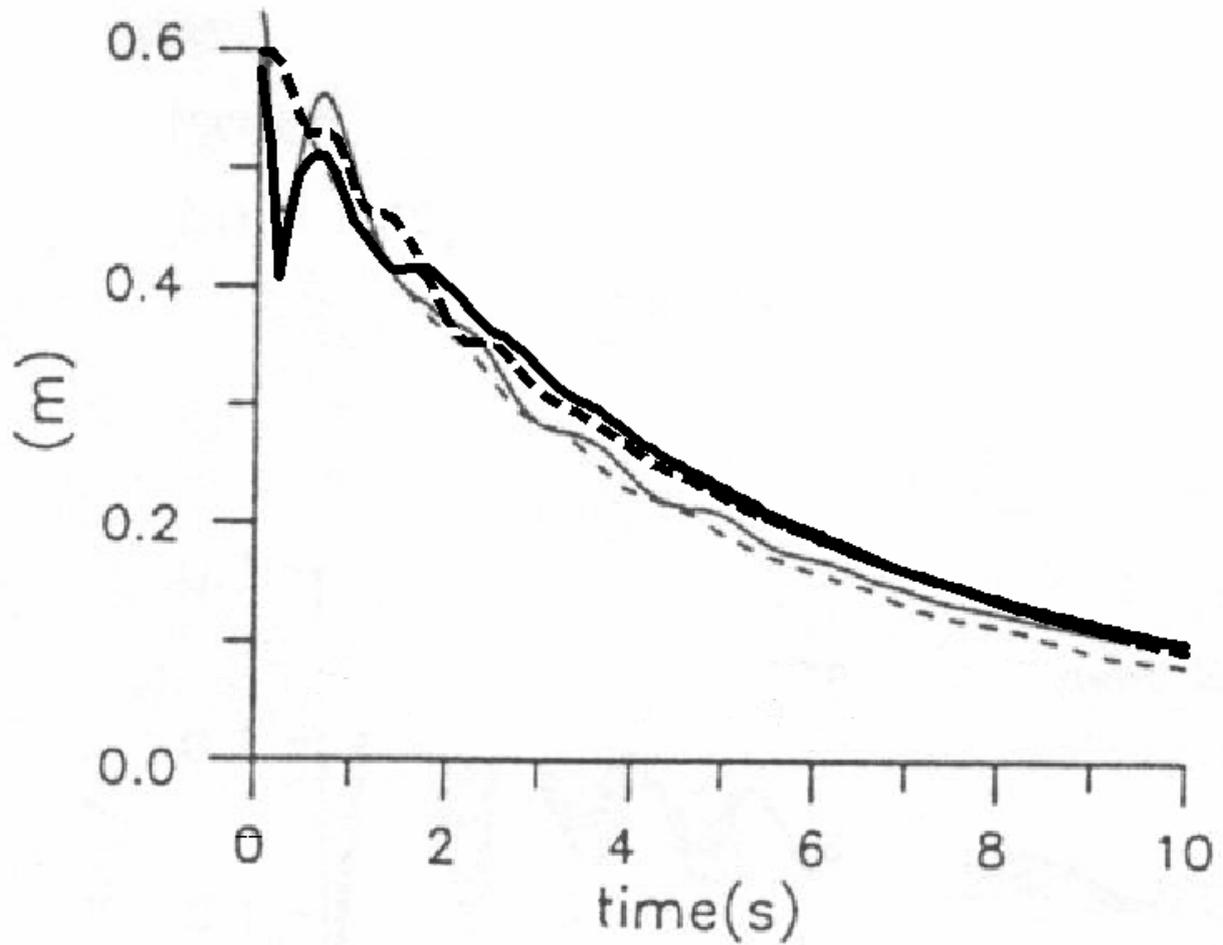

Fig. 23. Fluid depths on time for laboratory experiment with a sloping bed. Thin lines correspond to the data obtained in laboratory experiment: dashed – point P2, solid – point 0. Heavy lines correspond to numerical results of proposed quasi-two-layer method: dashed – point P2, solid – point 0

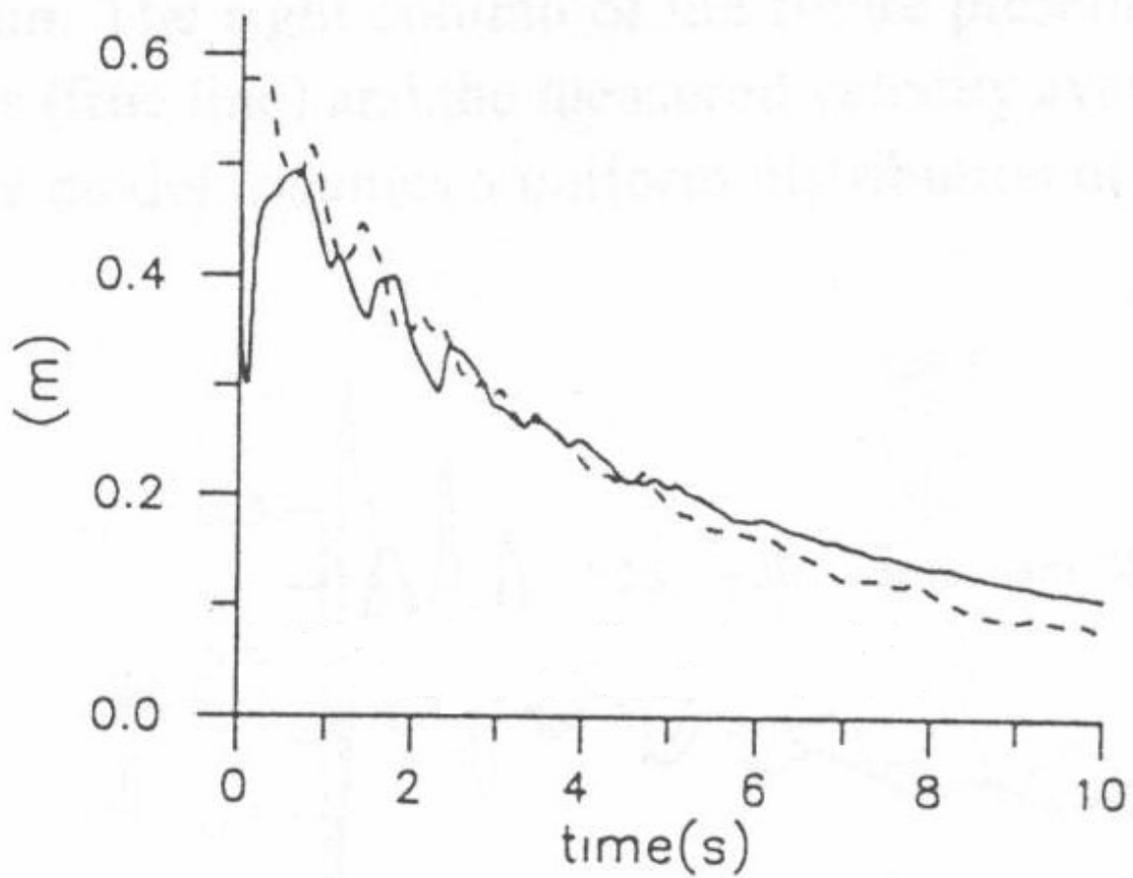

Fig. 24. Fluid depths on time for laboratory experiment with a sloping bed. Numerical results of WAF method: solid line – point 0, dashed line – point P2

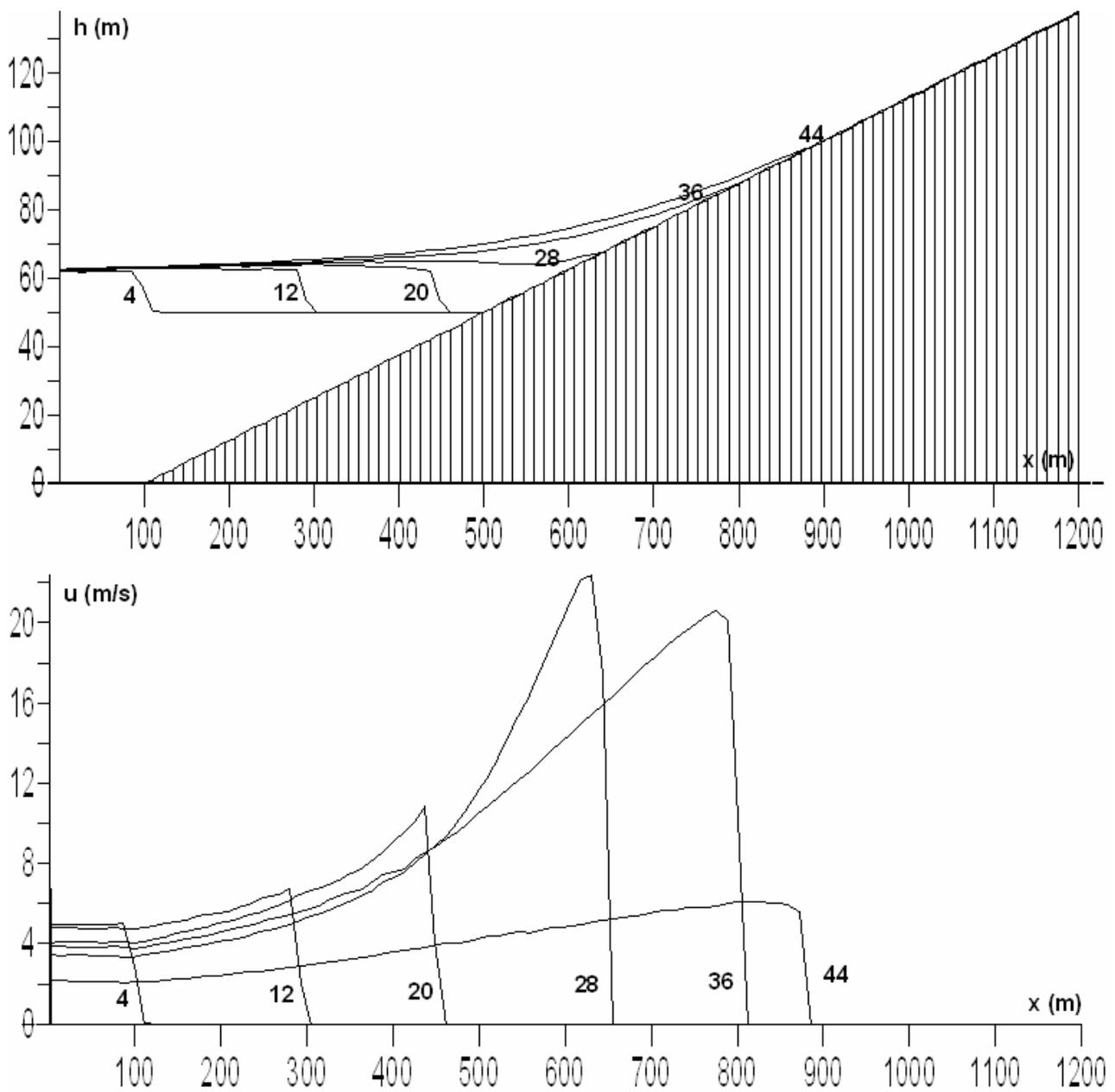
Fig. 25. Results of numerical computations of interaction of the Tsunami wave with the shore line

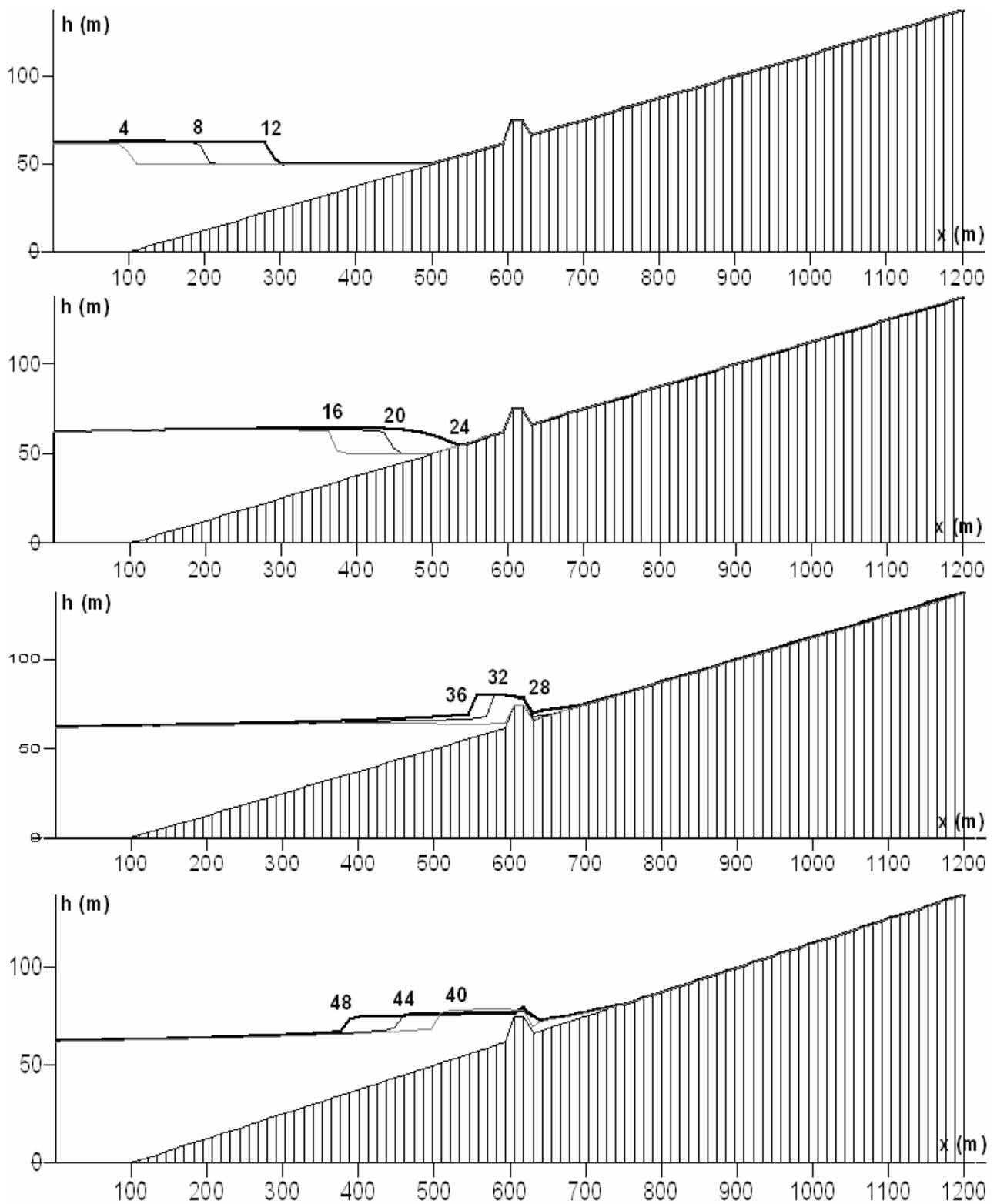

Fig. 26. Results of numerical computations of interaction of the Tsunami wave with the shore line with obstacle. Dynamics of the fluid depth in various time steps (indicated on the wave front in seconds)

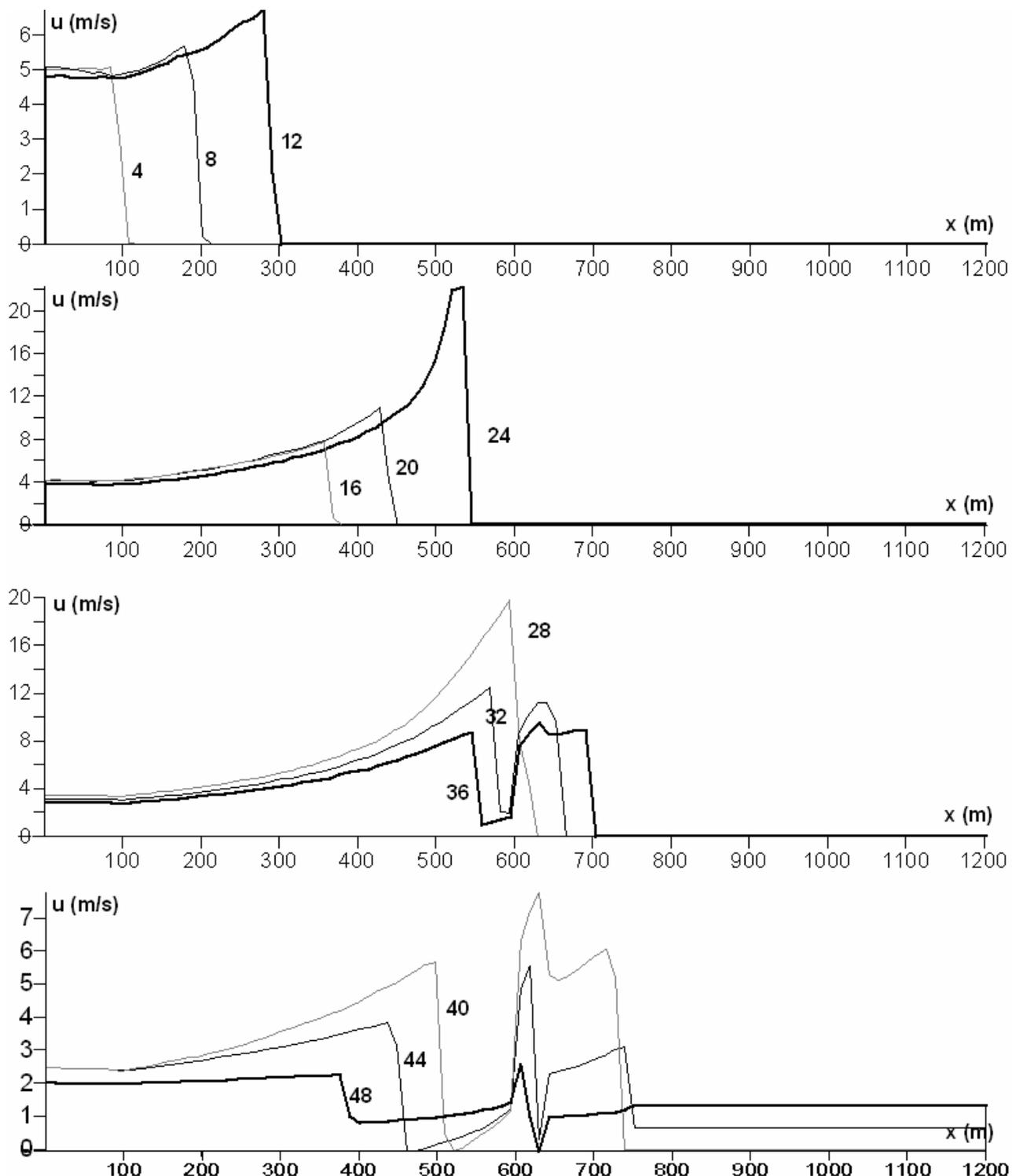
Fig. 27. Results of numerical computations of interaction of the Tsunami wave with the shore line. Dynamics of the fluid velocity in various time steps (indicated on the wave front in seconds)

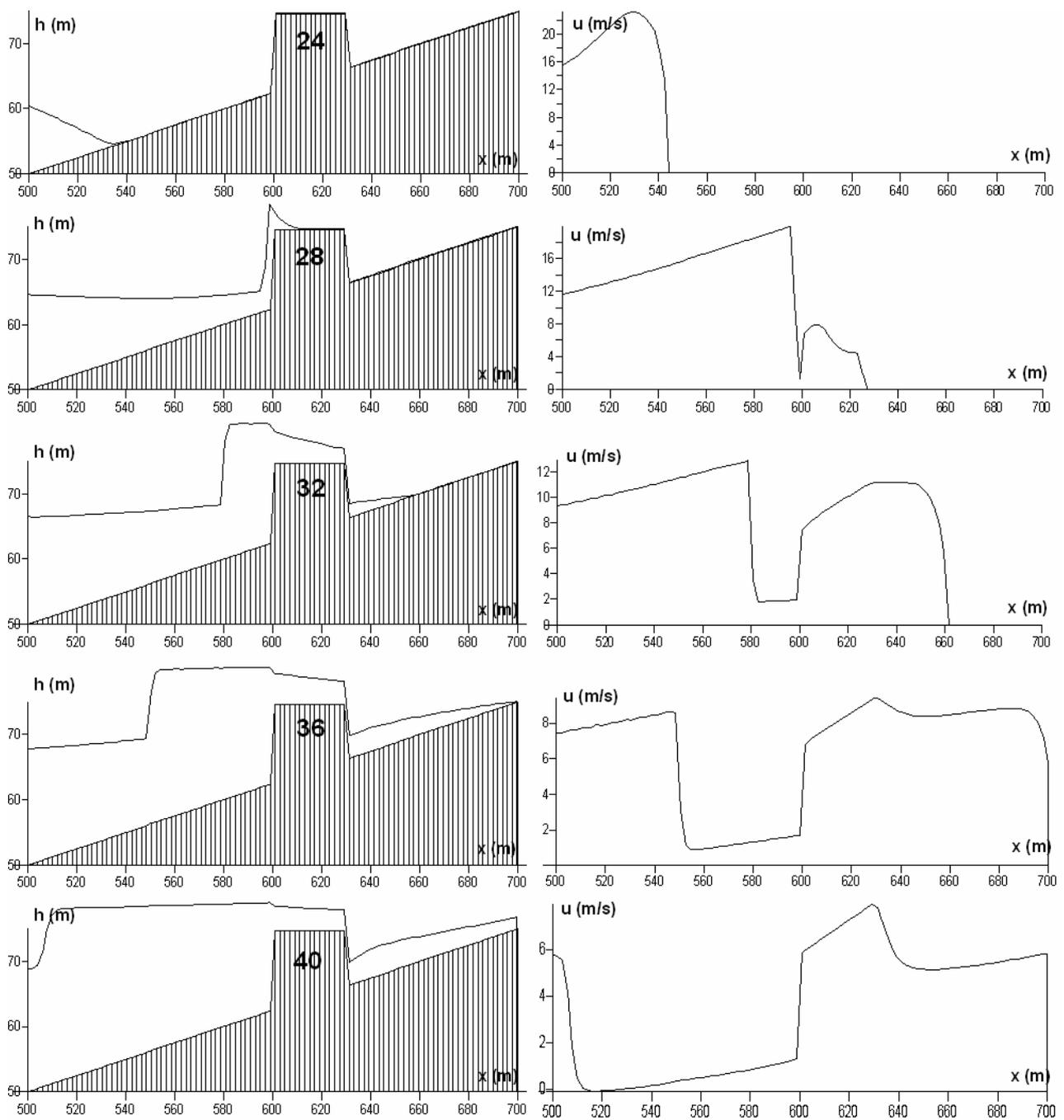

Fig. 28. Results of numerical computations of interaction of the Tsunami wave with the shore line with obstacles. Scaled-up area with an obstacle on the shore